\documentclass[a4paper,11pt]{article}
\pdfoutput=1
\usepackage{jheppub}

\usepackage{latexsym,amsbsy}
\usepackage{slashed}

\usepackage[compat=1.1.0]{tikz-feynman}

\newcommand{\BE}{\begin{equation} \begin{array}{c}}
\newcommand{\EE}{\end{array}\end{equation}}
\newcommand{\BEA}{\begin{equation} \begin{aligned}}
\newcommand{\EEA}{\end{aligned}\end{equation}}
\newcommand{\BT}{\begin{theorem}}
\newcommand{\ET}{\end{theorem}}

\newcommand{\bc}{\begin{center}}
\newcommand{\ec}{\end{center}}

\newcommand{\LX}{\Lambda}

\newcommand{\lX}{\lambda}

\newcommand{\ALG}{\mathcal{A}}

\newcommand{\VERMA}{\mathcal{V}}

\newcommand{\ZZ}{\mathcal{Z}}

\pgfkeys{/tikzfeynman/warn luatex=false}

\usepackage{pict2e}
\makeatletter\def\tikzfeynman@luatex@required@path{\relax}\makeatother
\makeatletter
\DeclareRobustCommand{\loplus}{\mathbin{\mathpalette\dog@lsemi{+}}}
\DeclareRobustCommand{\lotimes}{\mathbin{\mathpalette\dog@lsemi{\times}}}
\DeclareRobustCommand{\roplus}{\mathbin{\mathpalette\dog@rsemi{+}}}
\DeclareRobustCommand{\rotimes}{\mathbin{\mathpalette\dog@rsemi{\times}}}

\newcommand{\dog@rsemi}[2]{\dog@semi{#1}{#2}{-90,90}}
\newcommand{\dog@lsemi}[2]{\dog@semi{#1}{#2}{270,90}}
\newcommand{\dog@semi}[3]{%
  \begingroup
  \sbox\z@{$\m@th#1#2$}%
  \setlength{\unitlength}{\dimexpr\ht\z@+\dp\z@\relax}%
  \makebox[\wd\z@]{\raisebox{-\dp\z@}{%
    \begin{picture}(1,1)
    \linethickness{\variable@rule{#1}}
    \roundcap
    \put(0.5,0.5){\makebox(0,0){\raisebox{\dp\z@}{$\m@th#1#2$}}}
    \put(0.5,0.5){\arc[#3]{0.5}}
    \end{picture}%
  }}%
  \endgroup
}
\newcommand{\variable@rule}[1]{%
  \fontdimen8  
  \ifx#1\displaystyle\textfont3\else
    \ifx#1\textstyle\textfont3\else
      \ifx#1\scriptstyle\scriptfont3\else
        \scriptscriptfont3\relax
  \fi\fi\fi
}
\makeatother

\newcommand{\Prec}{{\prec}}
\newcommand{\Succ}{{\succ}}

\begin{document}

\title {
  Explicit construction of the finite dimensional indecomposable representations
  of the simple Lie-Kac $SU(2/1)$ superalgebra 
  and their low level non diagonal super Casimir operators.
}


\author{Jean Thierry-Mieg${}^1$, Peter Jarvis${}^{2,3}$ and Jerome Germoni${}^4$.}

\affiliation{${}^{1}$NCBI, National Library of Medicine, National Institute of Health, \\
  8600 Rockville Pike, Bethesda MD20894, U.S.A. \\
  ${}^{2}$School of Natural Sciences (Mathematics and Physics),\\
  University of Tasmania, Private Bag 37,\\
  Hobart, Tasmania 7001, Australia\\
  ${}^3$Alexander von Humboldt Fellow,
  ${}^4$Univ Lyon, Université Claude Bernard Lyon 1, CNRS UMR 5208, Institut Camille Jordan, F-69622 Villeurbanne, France.
  .}

\emailAdd{mieg@ncbi.nlm.nih.gov, peter.jarvis@utas.edu.au, germoni@math.univ-lyon1.fr}

\abstract{
  All finite dimensional irreducible representations of
  the simple Lie-Kac super algebra $SU(2/1)$ are explicitly constructed in
  the Chevalley basis as complex matrices. For typical representations, the distinguished Dynkin label is not quantized.
  We then construct the generic atypical indecomposable quivers
  classified by Marcu, Su and Germoni and 
  typical indecomposable $N$ generations block triangular extensions
  for any irreducible module and any integer $N$.
  In addition to the quadratic and cubic super-Casimir operators $C_2$ and $C_3$,
  the supercenter of the enveloping algebra
  contains a chiral ghost super-Casimir operator $T$ of mixed order (2,4)
  in the odd generators,
  proportional to the superidentity grading operator $\chi$,
  and satisfying $T = \chi\;C_2$ and we define a new factorizable chiral-Casimir
  $T^-=C_2(1-\chi)/2=(UV+WX)(VU+XW)$ where
  $(U,V,W,X)$ are the odd generators.. In most indecomposable
  cases, the super-Casimirs are non diagonal. We compute their pseudo-eigenvalues.
}

\maketitle

\section {Introduction.}
Recent progress on the $SU(2/1)$ model of the electroweak interaction \cite{TMJ21a}
motivates a thorough analysis of the finite dimensional representations of the $SU(2/1)$ superalgebra.

At first sight, this problem looks simple and one would think that the solution is well
known. Very early, Scheunert Nahm and Rittenberg \cite{SNR} have studied the finite dimensional
irreducible representations of $SU(2/1)$, and Kac has established \cite{Kac0,Kac1,Kac2} that for all the basic
classical Lie superalgebras $SU(m/n)$, $OSp(m/n)$, $D(2/1,\alpha)$, $G(3)$ and $F(4)$, one can choose
a set of positive roots, construct the Cartan matrix, classify the finite dimensional
irreducible representations by the Dynkin labels of their highest weight, and compute their
characters via an elegant generalization of the Weyl formula \cite{Kac3}.

However, the devil hides in the details. As shown by Kac, the representations can be
typical or atypical. The general Weyl formula only holds for the typical representations
which have vanishing superdimension. Many representations, in particular the fundamental
representation of nearly all the superalgebras listed above, are atypical: their
superdimension does not vanish, and the Weyl formula must be amended \cite{JHKTM90,KacWakimoto94,STMV17} or remains unknown.
Although tables of representations based on the ring structure
of the irreducible modules have been available for a long time \cite{TM83,TM85}, the
literature is abundant, and sometimes misleading.
For example, supertableaux methods, introduced by Dondi and Jarvis \cite{DondiJarvis81} and by Bars and Balantekin \cite{BalentekinBars81a,BalentekinBars81b}
are delicate requiring the introduction of composite
supertableaux \cite{DondiJarvis81,King83,Moens06}.
Another surprising property of superalgebras is the existence of indecomposable 
representations. In a semisimple Lie algebra, all finite dimensional representations are fully
reducible. But $SU(2/1)$ admits a rich zoo of finite dimensional indecomposable representations,
discussed for example by Marcu in 1980 \cite{Marcu80}, by Su in 1992 \cite{Su92,SuWang07}, and
classified by Germoni in 1997 \cite{Germoni97,Germoni98}\,.
Following the early work on $SU(2/1)$ as a model of electroweak interactions \cite{N1,F1,DJ1,NTM1}, Coquereaux in 1991 \cite{CQ91} applied these ideas about indecomposable representations to the particle
spectrum, and interpreted the representations of dimension 8 and 12, identified by Marcu, as describing the mixing 2 or 3 generations of leptons or quarks,
parameterized by the Cabibbo-Kobayashi-Maskawa mixing angles \cite{HS98,HPS98}.
Nevertheless, these representations were not considered in
several later studies, because other authors, for example G\"{o}tz, Quella and Schomerus \cite{GQS07},
restricted attention to the cases where the Cartan subalgebra is diagonalizable.

The classification of the Casimir operators, the generators of the supercenter of the enveloping algebra, is also much richer than originally anticipated. As shown by Arnaudon,
Bauer and Frappat \cite{ABF97}, a Lie-Kac super algebra can act on its enveloping algebra in the
standard way, but also in an alternative way, where the parity of the elements of the enveloping algebra is to a certain extent inverted. From this double action they deduced the
existence of and constructed a new kind of Casimir operator $T$ \cite{ABF97} , they call the ghost Casimir,
which acts as the square root of the super-Killing quadratic Casimir operator $C_2$ and is
proportional to the superidentity grading operator $\chi$ of the superalgebra.
This operator was also discovered by Musson {\cite{Musson97,Musson12} and generalized by Gorelik \cite{Gorelik01}.
It leads to interesting applications in super-geometry \cite{DufloPetracci07}
and is related to our recent definition of the parallel transport on a super-fiber bundle  \cite{TM20a}

In the present study, we provide an explicit construction of all
finite dimensional irreducible representations of $SU(2/1)$ in the Chevalley basis.
In this construction, matrix representatives of the even generators have integer entries,
and the odd generators have matrix elements of the form $(p + qb)/(a + 1)$ ,
where $(a, b)$ are the Dynkin labels and $(p,q)$ are integers.
Using this, we then show
how to combine atypical irreducible representations into indecomposable
quivers (decorated oriented graphs), illustrating in a concrete way the abstract classifications of Su and Germoni [17, 19].

Canonical examples are provided by the many different indecomposable decompositions
of the tensor product of two fundamental quartets, depending on their distinguished Dynkin labels.
Then we provide complete details of all variants of the constructions of Marcu \cite{Marcu80} and of Coquereaux \cite{CQ91}\,, and show that, in addition to the
doubling and the tripling of the fundamental quartet representation of $SU(2/1)$ demonstrated in the latter work, we can construct $N$ generations indecomposable block triangular representations $R(a,b,N )$,
for any choice of the Dynkin labels $(a,b)$ and any integer $N$.
To our knowledge, these $N$-generation representations are new.

Finally, we analyze the Casimir operators. In the atypical quivers they vanish.  In the $N$ generations indecomposable cases,
they are non-diagonal, but we can compute their pseudo eigenvalues and construct a multiple of the identity and of the superidentity
as a higher operators in the enveloping algebra. In all representations we compute the pseudo eigenvalues of
the chiral ghost Casimir $T$ (also called the Gorelik element) and of the classic quadratic super Casimir $C_2$ and show that
they are proportional: $T = \chi\;C_2$, i.e. $(T)^2 = (C_2)^2$. Since this equality holds in all finite dimensional representations,
it holds as an operator identity in the enveloping superalgebra. Furthermore, we introduce a new  chiral Casimir operator $T^- = C_2 (1-\chi)/2$
and show that it is factorizable $T^- = (UV + WX)(VU + XW)$, where $(U,V,W,X)$ are the four odd generators of $SU(2/1)$.

  Notice that we only refer to complex Lie algebras and Lie-Kac superalgebras, never to real forms, Lie groups or supergroups.
  Since there is no possible confusion, we use the standard physics notation, e.g. $SU(2/1)$ meaning $sl(2/1,C)$, to denote these algebras.

\section{Irreducible representations of low rank Lie superalgebras.}
\subsection {SU(2)}

To fix the notations, we recall in this section the well known construction of the finite
dimensional linear representations of $SU(2)$ over the complex field,
the only number field that we will consider.
In the Chevalley basis, the fundamental representation of the Lie algebra $A(1)=SU(2)$
for an arbitrary element $\LX$ is given by the matrices
\BE
F = \begin{pmatrix} 
  0 & 0 \cr
  1 & 0
 \end{pmatrix}
\,,\;\;
H = \begin{pmatrix} 
  1 & 0 \cr
  0 & -1
 \end{pmatrix}
\,,\;\;
E = \begin{pmatrix} 
  0 & 1 \cr
  0 & 0
\end{pmatrix}
\;
\Leftrightarrow\;\;
\LX = \begin{pmatrix} 
  h & e \cr
  f & -h
 \end{pmatrix}
\;.
\EE
The commutation relations are:
\BE
 [H,E]=2E\;,\;\;[H,F]=-2F\;,\;\;[E,F]=H\;.
 \EE
 
A highest weight representation with Dynkin label $(a)$
can be generated by defining a highest weight vector $|a>$ satisfying the conditions:
\BE
H|a> = a |a>\;,\;\;E|a> = 0\;.
\EE
and then considering the infinite vector space $\VERMA(a)$, called the Verma module,
generated by the free repeated action of $F$ on the highest weight vector $|a>$:
\BE
|a - 2p - 2 > = F |a-2p>\;,
\\
\VERMA(a) = \oplus |a - 2p>\;,\;\; p=0,1,...
\EE
Using the commutation relations (2.2) and the fact (2.3) that $E$ annihilates the highest
weight, we establish by recursion the action of $H$ and $E$ on all the elements of $\VERMA(a)$:
\BE
H|a-2p> = (a-2p) |a-2p>\;,\;\;
\\
E|a-2p>=p (a+1-p)|a-2p+2>\;.
\EE
If $a$ is an integer, we observe that the state $|-a-2>$ is the only other highest weight vector
in $\VERMA(a)$: $E|-a-2>=0$.
Quotienting out this submodule, we obtain a finite dimensional irreducible representation
\BE
R(a) = \VERMA(a)/\VERMA(-a-2)
\EE
of dimension $(a+1)$ and on this basis, all the matrix elements representing $(H,E,F)$
are integer numbers given by equations (2.4) and (2.5).
As an example, the 4 dimensional representation $R(3)$ is given in appendix A.
All finite dimensional representations of $SU(2)$ are
of this form, because the single Cartan operator $H$ can always be diagonalized and if the
representation is finite dimensional, it must contain a highest weight.

The non zero entries of the Killing metric are
\BE
g_{AB} = \frac{1}{2} \;Tr (\lX_A\lX_B)\;,\;\;
\\
g_{HH} = 2g_{EF} = \frac{\alpha}{6},\;\;\;g^{EF} = 2g^{HH} = \frac{12}{\alpha},
\EE
where $\alpha = a(a+1)(a+2)$.
In any irreducible representation, the Killing metric is proportional to the Killing metric
$\breve{g}$ of the fundamental 2-dimensional representation $(a=1)$
A direct calculation shows that the quadratic operator
\BE
C_2 = \frac{1}{2}\;\breve{g}^{AB}\;\lX_A\lX_B 
\EE
is a Casimir operator commuting with all the generators of the Lie algebra,
and proportional in $R(a)$ to the $(a + 1)$
dimensional identity matrix I, with eigenvalue $a(a + 2)/2$:
\BE
C_2 = \frac{1}{2}\;(HH + 2(EF + FE)) = \frac{a(a+2)}{2}\;I
\EE
In a reducible representation, the Casimirs are diagonal, with a possibly different eigenvalue
corresponding to each irreducible block.
In the following sections, we will apply a generalization of this method to the superalgebras
$OSp(1/2)$ and $SU(2/1)$.

\subsection {OSp(1/2)}

In the Chevalley basis, the fundamental representation of the superalgebra $B(1) = OSp(1/2)$ for an arbitrary element $\LX$ is given by the matrices

\BE
F = \begin{pmatrix} 
  0 & 0 & 0 \cr
  1 & 0 & 0 \cr
  0 & 0 & 0
 \end{pmatrix}
\;,\;\;
H = \begin{pmatrix} 
  1 & 0 & 0 \cr
  0 &-1 & 0 \cr
  0 & 0 & 0
 \end{pmatrix}
\;,\;\;
E = \begin{pmatrix} 
  0 & 1 & 0 \cr
  0 & 0 & 0 \cr
  0 & 0 & 0
\end{pmatrix}
\;,
\\
\\
R = \begin{pmatrix} 
  0 & 0 & 0 \cr
  0 & 0 & 1 \cr
 -1 & 0 & 0
 \end{pmatrix}
\;,\;\;
\chi = \begin{pmatrix} 
  1 & 0 & 0 \cr
  0 & 1 & 0 \cr
  0 & 0 & -1
\end{pmatrix}
\;,\;\;
S = \begin{pmatrix} 
  0 & 0 & 1 \cr
  0 & 0 & 0 \cr
  0 & 1 & 0
 \end{pmatrix}
\;
\\
\\
\Leftrightarrow
\LX = \begin{pmatrix} 
  \chi+h & e & s \cr
  f & \chi-h & t \cr
  -t & s & -\chi
 \end{pmatrix}
\;.

\EE
The generator $R$ corresponds to the single black odd root of the Dynkin diagram. The
commutators are :
\BE
   [H,E] = 2E\;,\;\;\;[H,F] = - 2F\;,\;\;[E,F] = H\;,
   \\
   \,[H,R] = R\;,\;\;[H,S] = -S \;,\;\;[E,S] = R\;,
   \\
   \,[E,R] = 0\;,\;\;[F,R] = S\;,\;\;[F,S] = 0\;.
\EE

The anticommutators are
\BE
\{S,S\} = -2 F\;,\;\;\{R,S\} = -H\;,\;\;\{R,R\} = 2E\;.
\EE
$(H,E,F)$ are even and generate $SU(2)=Sp(2)$, while $(R,S)$ are odd.
A highest weight is an eigenvector of $H$,  $H|a>=a|a>$ annihilated by the odd positive generator  $R$, $R|a>=0$.
A highest weight module can be understood as a superpolynome in $(F,S)$ acting on $|a>$  modulo the relations $SF =FS$ and $S^2 = -F$.
The irreducible highest weight representations are constructed by quotienting out the
invariant submodules. Since
\BE
R|a>=0\; =>\\
\,E|a> = RR |a> = 0\;,\;\; ES|a> = [E,S]|a> = R|a> = 0,
\\
R F^p |a> = -p F^{p-1}S |a>\;,\;\;R F^p S|a> = (2p-a) F^p |a>\;,
\\
EF^p|a> = p (a + 1 - p) |a>\;,\;\;EF^pS|a> = p (a - p) S|a>,
\EE
 the irreducible module is finite dimensional if and only if $a$ is an integer and consists of two
 finite $Sp(2)$ modules with highest weights $|a>$ and $|a-1> = S|a>$, except in the trivial case $a=0$.
The matrices $\lX_A$ of all the complex finite dimensional irreducible representations of  $OSp(1/2)$ are
given in appendix B. They only involve signed integers.
 
The grading operator $\chi$, called here the chirality or the superidentity, is not a generator of the
superalgebra but is defined in all its representations. It commutes with the even operators and anticommutes
 with the odd operators
\BE
 \chi^2 = 1\;,
 \\
 \,[\chi,H] = [\chi,E] = [\chi,F] = 0\;,
 \\
 \, \{\chi,R\} = \{\chi,S\} = 0\;.
 \EE

 In a representation space, the chirality of the highest weight can be chosen at will as $\pm 1$.
 The supertrace is defined in terms of the chirality as
\BE
 STr(M) = Tr (\chi M)\;.
\EE
Selecting $+1$ on the highest weight, the non zero entries of the super-Killing metric are
\BE
g_{AB} = \frac{1}{2} \;STr (\lX_A\lX_B)\;,\;\;g^{AB}g_{BC} = \delta^A_{\;\;C}
\\
g_{HH} = a(a+1)/2,\;\;g_{EF}=g_{FE}=a(a+1)/4,\;\;g_{RS}=-g_{SR} = a(a+1)/2\;.
\\
g^{HH} = 2/a(a+1),\;\;g^{EF}=g^{FE}= 4/a(a+1),\;\;g^{RS}=-g^{SR} = -2/a(a+1)\;.

\EE
The quadratic Casimir operator (beware of the order in which the odd indices are contracted) defined
using the fundamental triplet $a=1$ Killing metric $\breve{g}$ reads
\BE
C_2 = \frac{1}{2}\breve{g}^{BA}\;\lX_A\lX_B = \frac{1}{2}(HH + 2(EF + FE) + RS - SR) = \frac{a(a+1)}{2}\;I\;.
\EE
As discovered by Arnaudon, Bauer and Frappat in 1997 \cite{ABF97} and Musson {\cite{Musson97}, we can construct a
quadratic chiral-Casimir operator, referred to by these authors as the ghost Casimir operator, proportional to the super identity grading operator $\chi$:
\BE
  T =  [S,R] + \frac{1}{2}\;I = (a + \frac {1}{2})\;\chi\;,
\EE
and we have the curious identity
\BE
[S,R] + ([S,R])^2 = 2 \;C_2\;.
\EE

\subsection {SU(2/1)}

In the Chevalley basis, the fundamental representation of the Lie superalgebra
$SU(2/1)$ for an arbitrary element $\LX$ is given by the matrices

\BE
F = \begin{pmatrix} 
  0 & 0 & 0 \cr
  1 & 0 & 0 \cr
  0 & 0 & 0
 \end{pmatrix}
\;,\;\;
H = \begin{pmatrix} 
  1 & 0 & 0 \cr
  0 &-1 & 0 \cr
  0 & 0 & 0
 \end{pmatrix}
\;,\;\;
E = \begin{pmatrix} 
  0 & 1 & 0 \cr
  0 & 0 & 0 \cr
  0 & 0 & 0
\end{pmatrix}
\;,
\\
\\
Y = \begin{pmatrix} 
 -1 & 0 & 0 \cr
  0 &-1 & 0 \cr
  0 & 0 &-2
 \end{pmatrix}
\;,\;\;
\chi = \begin{pmatrix} 
  +1 & 0 & 0 \cr
  0 & +1 & 0 \cr
  0 & 0 & -1
\end{pmatrix}
\\
\\
U = \begin{pmatrix} 
  0 & 0 & 0 \cr
  0 & 0 & -1 \cr
  0 & 0 & 0
 \end{pmatrix}
\;,\;\;
V = \begin{pmatrix} 
  0 & 0 & 0 \cr
  0 & 0 & 0 \cr
  0 & 1 & 0
 \end{pmatrix}
\;,\;\;
\\
\\
W = \begin{pmatrix} 
  0 & 0 & -1 \cr
  0 & 0 & 0 \cr
  0 & 0 & 0
 \end{pmatrix}
\;,\;\;
X = \begin{pmatrix} 
  0 & 0 & 0 \cr
  0 & 0 & 0 \cr
  1 & 0 & 0
 \end{pmatrix}
\;.
\\
\\
\Leftrightarrow
\LX = \begin{pmatrix} 
  \chi+h-y & e & -w \cr
  f & \chi-h-y & -u \cr
  x & v & -\chi -2y
 \end{pmatrix}
\;.
\EE
The 3 even generators, $(F,H,E)$ generate $SU(2)$, compare with (2.1). The hyper-charge $Y$
commutes with $(F,H,E)$ and generates $U(1)$. The 4 odd
generators $(U,V,W,X)$ form two $SU(2)$ doublets $(U,W)$ and $(X,V)$

\BE
[H,U]=-U,\;[H,W]=W,\;\;\;[H,X]=-X,\;[H,V]=V,
\\
\,[Y,U]= U,\;[Y,V]=-V,\;\;\;[Y,W]= W,\;[Y,X]=-X,
\\
\,[E,U]=W,\;\;\;[F,W]=U,\;\;\;[E,X]= -V,\;\;\;[F,V]=-X\;,\;\;
\\
\,[E,V]=[E,W]=[F,U]=[F,X]=0,
\EE
and satisfy the anticommutation relations
\BE
\{U,V\} = K = (Y + H)/2,\;\;\;\{W,X\}= (Y-H)/2,\;\;\;
\\
\{U,X\} = -F,\;\;\;\{V,W\} = -E,\;\;\;\{U,W\} = \{X,V\} = 0\;,
\\ 
\; [K,E] = E\;,\;\;[K,F] = -F\;,\;\;[K,U] = [K,V] = 0\;.
\EE

The mapping $(F,H,E) \rightarrow (F,H,E)$, $Y \rightarrow -Y$, $(U,V,W,X) \rightarrow (W,X,U,V)$ defines an automorphism of the superalgebra \cite{SNR}.

The superidentity, or chirality, operator $\chi$ is not a generator of the superalgebra but defines the grading.
$\chi$ commutes with all the even generators
and anticommutes with all the odd ones and defines the supertrace $STr(M) = Tr (\chi M)$.

Comparing the matrices (2.10) and (2.20), an embedding of $OSp(1/2)$ in $SU(2/1)$ is
defined by the relations

\BE
H = H\;,\;\; E = E\;,\;\; F = F\;,\;\; R = V - W\;,\;\; S = -U-X
\EE

The quotient space $(U-X,Y,V+W)$ forms
the fundamental triplet of $OSp(1/2)$.

The finite dimensional highest weight representations $V(a,b)$ of $SU(2/1)$
are characterized by their highest weight vector $|\LX>=|a,b>$, where $(a)$ is
a non negative integer, and $(b)$ a complex number
\BE
  H|\LX>=a |\LX>\;,\;\;K|\LX>=b|\LX>\;,\;\;E|\LX>=U|\LX>=0\;.
\EE
They contain four
$SU(2)$ submodules  with  weights
\BE
|a-2n,b-n> = F^n|\LX>,\;n=0,1,2,...,
\\
|a+1-2n,b-n> = F^nV|\LX>,\;n=0,1,2,...
\\
|\omega> = (aFV-(a+1)VF))|\LX>\;,\;\;
\\
|a-1-2n,b-1-n> = F^n|\omega>,\;n=0,1,2,...,
\\
|a-2n,b-n-2> = F^n VFV |\LX>,\;n=0,1,2,....
\EE
labeled by the eigenvalues of $(H,K)$. For any vector $|a',b'>$ in these sets
\BE
H |a',b'> = a' |a',b'>\;,\;\;Y|a',b'> = (2b'-a')|a',b'>\;,\;\;\\K|a',b'>=b'|a',b'>\;.
\EE
The action of $F$ is explicit $F|a,b'> = |a'-2,b'-1>$.
We then check that $|\omega>$ is an $SU(2)$ highest weight $E|\omega>=0$.
Finally, the action of $E$
can be computed in the 4 cases by pushing $E$ all the way to the right and using
in the fourth case $VV=0$:
\BE
E|a-2n,b'-n>=E F^n|\LX>
\\
=\sum F^i [E,F]F^{n-i-1}|a,b'>=n(a-n+1)|a-2n+2,b'+1-n>\;.
\EE
We have $EF^{a+1}|a,b'> = 0$ which justifies that we can quotient out
from the $SU(2)$ Verma module $\VERMA(a) = \oplus F^n|a>, n=0,1,2,....$, the Verma
submodule $\VERMA(-a-2)$
and recover a finite dimensional representation $R(a)$ if $(a)$ is a non negative integer.
This also applies to the 3 other even submodules with $SU(2)$ highest weights $(V|\LX>,|\omega>,VFV|\LX>)$  
Notice that if $(a=0)$ , the $|\omega>$ submodule can also be quotiented out.

We now check the action of the odd raising operator $U$ on the four  $SU(2)$ highest weights. We have
\BE
U|\LX>= 0\;,\;\,
\\
UV|\LX>=b|\LX>\;,\;\;
\\
U|\omega>=(a+1-b)F|\LX>\;,\;\;
\\
UFVF|\LX>=(\frac{b}{a+1}-1)FV|\LX> + \frac{b}{a+1}|\omega>
\EE
There are two special cases, called respectively atypical 1 and atypical 2.
If $(b=0)$, the vector $V|\LX>$ is an $SU(2/1)$ highest weight
and its submodule can be quotiented out.
Similarly,
If $(b = a + 1)$, $|\omega>$ is an $SU(2/1)$ highest weight and its submodule can be quotiented out.
In both cases the $SU(2)$ submodule with highest weight $V F V |\LX>$ also disappears. This
shows that the atypical modules are nested: the state $V F V |\omega>$ is not part of the $SU(2/1)$
module with highest weight $|\LX>$, hence it cannot be part of the $SU(2/1)$ module
with highest weight $|\omega>$, similarly for the $SU(2/1)$ module with highest weight $V|\LX>$.
Hence these modules are also atypical. As a result, the super Casimir operators which have
the same eigenvalues on all states of the reducible Verma module have the same value on
successive members of these atypical chains. But since these chains respectively start and
end on the trivial representation $|0, 0 >$, the Casimir operators identically vanish on all
irreducible atypical representations. This property will be checked by inspection below.

We now compute the action of $U$ on all states using the fact the $U$ commutes with $F$ and that $V^2=0$.
\BE
UF^n|\LX>=0\;,\;\;
\\
UF^nV|\LX>=bF^n|\LX>\;,\;\;
\\
UF^n|\omega>=(a+1-b)F^{n+1}|a,b>\;,\;\;
\\
UF^nVFV|\LX>=(\frac{b}{a+1}-1)F^{n+1}V|\LX>+\frac{b}{a+1}F^n|\omega>
\;.
\;.
\EE
Finally, we compute the action of $(V)$
\BE
VF^n|\LX>=(1-\frac{n}{a+1})F^n|\LX> - \frac{n}{a+1}|\omega>\;,
\\
VF^nV|\LX>=nF^{n-1}VFV|\LX>\;,
\\
VF^n|\omega>=(a-n)F^nVFV|\LX>\;,
\\
VF^nVFV|\LX>=0\;.
\EE
The remaining odd operators operators $(W,X)$ are deduced using the relations
$[E,U]=W$ and $[F,V]=X$.

In this way, we have completed the construction of all the irreducible finite dimensional
representations $R(a,b)$ of $SU(2/1)$ in terms of even matrices values in $Z$
and odd matrices with entries in $(p+qb)/(a+1)$, with $(p,q)$ also in $Z$.

Examples of typical and atypical representations are given in the appendices C, D, E.

\section{Indecomposable representations: examples.}
\subsection {The indecomposable atypical Kac modules.}

In the $SU(2)$ Lie algebra case, the Verma module is infinite dimensional, so it is imperative
to pass to the quotient to generate a finite dimensional representation. But in a Lie
superalgebra, this is not necessary. Starting from an atypical 1 representation with highest
weight $|\LX>$, we do not need to quotient out the secondary highest weight $V|\LX>$, but we can
scale it arbitrarily by a non zero factor $\alpha$ and construct a one-parameter indecomposable
representation. For example, the adjoint representation of $SU(2/1)$ is 8 dimensional and
its Dynkin labels are $(a=1,b=1,y=1)$. If we now consider the shifted adjoint representation (appendix E)
with Dynkin labels $(a=1,b=0,y=-1)$, it is naively also 8 dimensional, but it satisfies $(b = 0)$ and is
atypical 1. Its irreducible component is of dimension 3 and corresponds to the fundamental
representation given by the matrices (24). The state $V|1, 0>$ is a highest weight. Rather
than setting it to zero, we define $V |1, 0 >= \alpha|2,0>$. The state $|2,0>$ is the highest weight
of a secondary 5 dimensional atypical 1 representation of $SU(2/1)$ and we have constructed
in this way a one-parameter family of 8 dimensional indecomposable representations of
$SU(2/1)$ where all the entries in the $3\times 5$ lower left corner are proportional to $\alpha$.
This indecomposable module is denoted
$8^\Succ_{-1} = 3_{-1} \Succ 5_{-2}$.
The
matrices are given explicitly in appendix G.

The general aspect of the highest weight indecomposable matrix has 4 blocks
\BE
M^{\Succ} = \begin{pmatrix}
  A & 0 \cr
  \alpha & B 
\end{pmatrix}
\EE
where $A$ is the highest weight irreducible block, $B$ the block that can be quotiented out, and $\alpha$
the transition block. Alternatively we can construct an indecomposable representation from the orbit of the
lowest weight by keeping the other transition block
\BE
M^{\Prec} = \begin{pmatrix}
  A & \beta \cr
  0 & B 
\end{pmatrix}
\EE
Consider the highest weight vector $(a=3,b=4,Y=5)$. The typical representation with $(a=3)$ splits, with respect to the even algebra as $(4 ; 3 + 5 ; 4)$.
But since $b=a+1=4$, this representation is atypical 2 and splits as $9_5=(a=3,b=4,Y=5;a=4,Y=4)$ plus $7_4=(a=2,b=3,Y=4;a=3,Y=3)$. The highest weight indecomposable
is $16^{\Succ}_5 = 9_5 \Succ 7_4$. The lowest weight indecomposable is $16^{\Prec}_5= 9_5 \Prec 7_4$. We now consider the representation with highest weight $(a=3,b=0,Y=-3)$.
It is atypical 1 and split as $7_{-3}=(a=3,b=0,Y=-3;a=2,Y=-4)$ plus $9_{-4}=(a=4,b=0,Y=-4;a=3,Y=-5)$. We can now construct a highest weight
indecomposable $16^{\Succ}_{-3}=7_{-3} \Succ 9_{-4}$ and a lowest weight indecomposable $16^{\Prec}_{-3}=7_{-3} \Prec 9_{-4}$. The l.w. indecomposable
$7_{-3} \Prec 9_{-4}$ is the anti-representation of
the h.w. indecomposable $9_4 \Succ 7_3$ and vice versa.

These constructions are valid for all atypical representations. In each case, the atypical irreducible representation can be seen as the
special $\alpha=0$ case of a 1 parameter $\alpha$-family of indecomposable representations.

Reciprocally, since the odd generators form $SU(2)$ doublets with $Y=\pm 1 $, only successive irreducible atypical matrices, such that the
$Y$ charge of their highest weight differ by at most one unit can be connected in this way, and they always belong to the same Kac module.

Alternatively, let us now try to intertwine two identical atypical representations $A$ with the same Dynkin labels $(a,b)$. The matrix $R(A)$, as an even module, can be written
\BE
R = \begin{pmatrix}
 M  & \alpha \cr
  \beta & N 
\end{pmatrix}
\EE
where $(M,N)$ are $SU(2)$ modules with Dynkin labels $(a,a+1)$ if $A$ is atypical 2 $(b=a+1)$. The atypical 1 case $(b=0)$ is equivalent.
The contact matrices $(\alpha,\beta)$ are uniquely defined knowing $(a)$. The candidate indecomposable odd matrices $S_1$ will therefore be of the form
\BE
S_1 = \begin{pmatrix}
 0  & \alpha & 0 & 0\cr
 \beta & 0 & 0 & 0\cr
 0 & s  \alpha & 0 & \alpha\cr
 t \beta & 0 & \beta & 0
\end{pmatrix}
\EE
where $(s,t)$ are complex parameters. But when we check the commutation relations $\{V,W\}=-E$ (2.22) we find $s+t=0$ and in this case
the representation can be block diagonalized by a change of variable affecting just the two copies of the $(a+1)$ module.

Conclusion: pairs of atypical irreducible representations of $SU(2/1)$ can be joined in an indecomposable module if and
only if they belong in a single Kac module. Therefore these indecomposable modules are smooth continuations of the
corresponding typical module.

\subsection {The indecomposable square of the fundamental quartet. }

To gain insight into the structure of larger indecomposable modules, we construct explicitly in appendix F
the tensor product two quartets with highest weights $(0,b)$ and $(0,b')$. Knowing all the generators of the quartets, it is immediate to construct
the orbits of the highest weight and of the lowest weight of the tensor product. We obtain 2 quartets, one with highest weight $(a=0,Y=2(b+b'))$
and one with lowest weight $(a=0,Y=2(b+b'-2))$. In general, they are irreducible and their complement in the tensor product is a shifted
adjoint with highest weight $(a=1,Y=2(b+b')-1)$. But if $b+b'$ equals $0$, $1$ or $2$ many cases must be distinguished.
As above, the ${}^\Prec$ and ${}^\Succ$ symbols indicates an indecomposable
representation and
the lower index indicates the $Y$ charge of the highest weight of the representation. $8_1$ is the
adjoint representation. The details of the calculations are presented in appendix F.

\begin{enumerate}

\item
  $b+b' = 1, bb' \neq 0$, the quartets are irreducible and conjugated. This case includes the square of the fundamental
  quartet $b=b'=1/2$ of $OSp(2/2)$. The tensor product decomposes as two very different octets:
  \BE
 (4_1)^2 = 8_1 \oplus (8_2^{\Succ \oplus \Succ} = 1_0 \Succ (3_2 \oplus 3_{-1})  \Succ 1'_0),
  \EE
  The $8_1$ is the irreducible adjoint representation of $SU(2/1)$. The other octet forms an indecomposable doubly connected  (or cyclic) graph of irreducible components.
\item    
  else if $b=1, b'=0$, we start with 2 indecomposable quartets, we must distinguish 4 cases:
  \BE
  (3_2 \Succ 1_0) \otimes (1_0 \Succ 3_{-1}) = 8_1 \oplus (4^{\Succ}_2 = 3_2 \Succ 1_0) \oplus (4^{\Succ}_0=1'_0 \Succ 3_{-1})
  \;,
  \\
  (3_2 \Succ 1_0) \otimes (1_0 \Prec 3_{-1}) = 8_1 \oplus  (8^{\Succ \oplus \Succ}_2 = 1_0\Succ (3_2 \oplus 3_{-1}) \Succ 1'_0)
  \;,
  \\
  \\
  (3_2 \Prec 1_0) \otimes (1_0 \Succ 3_{-1}) = 8_1 \oplus  (8^{\Succ \oplus \Succ}_2 = 1_0\Succ (3_2 \oplus 3_{-1}) \Succ 1'_0
  \;,
  \\
  (3_2 \Prec 1_0) \otimes (1_0 \Prec 3_{-1}) = 8_1 \oplus (4^{\Prec}_{2} = 3_{2} \Prec 1_0) \oplus (4^{\Prec}_0 = 1'_0 \Prec 3_{-1})

  \EE
\item  else if $b+b'=2, b \neq 1$, we get a different cyclic representation
  \BE
  (4_{2b}) \otimes (4_{2(2-b)}) = 4_4 \oplus (12^{\Succ\oplus\Succ}_2=3_{2}\Succ (5_{3}\oplus 1_0) \Succ 3'_{2})
  \EE
\item  else if $b+b'=0, b \neq 0$, we get the conjugated result:
  \BE
  (4_{2b}) \otimes (4_{-2b)}) = 4_{-2} \oplus (12^{\Succ\oplus\Succ}_1=3_{-1}\Succ(5_{-2}\oplus 1_0) \Succ 3'_{-1})
  \EE  
\item
  else if $b=b'=1$, we start again with 2 indecomposable quartets:
  \BE
  (3_2 \Prec 1_0) \otimes (3_2 \Prec 1_0) = 4_4 \oplus (4^{\Prec}_2=3_2 \Prec 1_0) \oplus (8^{\Prec}_3=5_3 \Prec 3'_2)
  \\
  (3_2 \Succ 1_0) \otimes (3_2 \Prec 1_0) = 4_4 \oplus (12^{\Succ\oplus\Succ}_2 = 3_2 \Succ (5_3 \oplus 1_0) \Succ 3'_2)
  \\
  (3_2 \Succ 1_0) \otimes (3_2 \Succ 1_0) = 4_4 \oplus (4^{\Succ}_2=3_2 \Succ 1_0) \oplus (8^{\Succ}_3=5_3 \Succ 3'_2)
  \EE
\item  else if $b=b'=0$, we get the conjugated result:
  \BE
  \\
  (1_0 \Prec 3_{-1}) \otimes (1_0 \Prec 3_{-1}) = 4_{-2} \oplus (4^{\Prec}_0 = 1_0 \Prec 3_{-1}) \oplus (8^{\Prec}_{-1} = 3'_{-1} \Prec 5_{-2})
  \\
  (1_0 \Succ 3_{-1}) \otimes (1_0 \Prec 3_{-1}) = 4_{-2} \oplus (12^{\Succ\oplus\Succ}_{-1} = 3_{-1} \Succ (5_{-2} \oplus 1_0) \Succ 3'_{-1})
  \\
  (1_0 \Succ 3_{-1}) \otimes (1_0 \Succ 3_{-1}) = 4_{-2} \oplus (4^{\Succ}_0 = 1_0 \Succ 3_{-1}) \oplus (8^{\Succ}_{-1} = 3'_{-1} \Succ 5_{-2})
  \EE
\item
  else, generic typical case:
  \BE
    (4_{2b}) \otimes (4_{2b'}) = 4_{2(b+b')}  \oplus 4_{2(b+b')-2} \oplus 8_{2(b+b')-1}
  \EE
\end{enumerate}

The most interesting point is the apparition of doubly connected (cyclic if we drop the orientation of the arcs of the quiver)
indecomposable representations
as described in Marcu \cite{Marcu80}, Germoni \cite{Germoni97}, Benamor \cite{Benamor97} or
G\"{o}tz, Quella and  Schomerus,  \cite{GQS07}.
These results are compatible with \cite{GQS07} but more detailed, as previous studies did not
distinguish the orientations like $4^{\Succ}_3$ and $4^{\Prec}_3$.
Earlier on, Su \cite{Su92} erroneously rejected the possibility of having a cycle.

\section{Indecomposable representations: general case.}
\subsection {Atypical indecomposable quivers.}

Extending on the analysis of the tensor product of two quartets, we can in general
chain 3 successive atypical modules $(A,B,C)$ as $A\Prec B  \Succ C$, with explicit matrix form:
\BE
M_B = \begin{pmatrix}
  A & \alpha & 0 \cr
  0 & B & 0 \cr
  0 & \beta & C \cr
\end{pmatrix}
\EE
For example if $(A,B)$ are the representations of dimension $(9,7)$ described below equation (3.2),
$C$ would be the representation of dimension $5=(a=1,b=2,Y=3;a=2,Y=2)$ and the transition matrices are copied from the relevant typical representations
and scaled by the complex numbers $(\alpha,\beta)$.
$M_2$ is a double mousetrap indecomposable representation $9 \Prec 7 \Succ 5$, and only the orbit of the central component covers the whole representation.
Alternatively we can construct the representation $9 \Succ 7 \Prec 5$
\BE
M_c = \begin{pmatrix}
  A & 0 & 0 \cr
  \alpha & B & \beta \cr
  0 & 0 & C \cr
\end{pmatrix}
\EE

These explicit constructions agree with the results of Su \cite{Su92} and Germoni \cite {Germoni97}
who proved the existence of these indecomposable triplets.

The next natural idea is to considered a 3-chain $A \Succ B \Succ C$:
\BE
M_c = \begin{pmatrix}
  A & 0 & 0 \cr
  \alpha & B & 0 \cr
  0 & \beta & C \cr
\end{pmatrix}
\EE
where $(A,B,C)$ are again successive atypical representations, for example $11 \Succ 9 \Succ 7$ or $11 \Succ 9 \Succ 11$. As before, the transition blocks are chosen
to coincide with the corresponding odd block of the Kac module (the would be typical representation). But we have to check if the square of
the odd generators vanishes, in particular the rectangular $3\times 7$ matrix matrix $\beta\alpha$ must vanish.
But we know the explicit form of these matrices (F.2), as they are constrained by the fact that the odd generators form an $SU(2)$ doublet.
We have:
\BE
\alpha_V = \begin{pmatrix}
  0 & -\alpha/3 & 0 & 0 & 0 & 0 & 0 \cr
  0 & 0 & -2\alpha/3 & 0 & 0 & 0 & 0 \cr
  0 & 0 & 0 & 0 & \alpha' & 0 & 0 \cr
  0 & 0 & 0 & 0 & 0 & 2\alpha' & 0 \cr
  0 & 0 & 0 & 0 & 0 & 0 & 3\alpha' \cr
\end{pmatrix}
\;,\;\;\;
\\
\\
\beta_V = \begin{pmatrix}
  0 &-\beta/2 & 0 & 0 & 0 \cr
  0 &   0 & 0 & \beta' & 0 \cr
  0 &   0 & 0 & 0 & 2\beta' \cr
\end{pmatrix}
\EE
The product $\beta_V\alpha_V$ does not vanish unless $\alpha\beta = \alpha'\beta' = 0$.
Try $\alpha' = \beta = 0$. But now at the position of the matrix $\alpha$, the anticommutator
$\{U,V\}$ no longer vanishes as it should unless $\alpha=0$ because it is not compensated by the $\alpha'$ term.
Recall that in the typical module, we would have in these notations $\alpha = \alpha'$.
The only possibility is to have a multiple of the identity in the lower left blocks.
This requires that $A$ and $C$ have the same dimension.
We conclude that an indecomposable 3-chain of type  $A \Succ B \Succ C$ is forbidden
but $A \Succ B \Succ A'$ is allowed, with $(A,B,C,A')$ atypical, and $dim(A)=dim(A') \neq dim(C)$. 

A contrario, let us now show that a representation of the type $A \Prec B \Succ A'$ is
decomposable.
Consider the matrix
\BE
M = \begin{pmatrix}
  A & \alpha & 0 \cr
  0 & B & 0 \cr
  0 & \beta & A \cr
\end{pmatrix}
\EE
by $SU(2)$ invariance, the contact matrices $(\alpha,\beta)$ are identical up to a scale factor that can be eliminated by rescaling
the carrier space. Then the resulting matrix can be decomposed using the change of coordinates:
\BE
U = \begin{pmatrix}
  1 & 0 & 1 \cr
  0 & 1 & 0 \cr
  -1 & 0 & 1 
\end{pmatrix}
\;\;\Rightarrow\;\;
U M U^{-1} = \begin{pmatrix}
  A & \alpha & 0 \cr
  0 & B & 0 \cr
  0 & 0 & A 
\end{pmatrix}
\EE

The non existence of triplets $A \Succ B \Succ C$ and $A \Prec B \Succ A'$ leaves the trivial possibility of
longer alternated chains of the type $A \leftrightarrow B \leftrightarrow C \leftrightarrow D \leftrightarrow E \leftrightarrow D ... $ where $(A,B,C,D,E)$ are successive atypical irreducible modules, for example $Y=(5,4,3,4,3,0,-1,0)$ (notice that $(Y=1)$ does not exist as the highest $Y$ weight of an
irreducible atypical module, so (0,2) are counted as neighbors)
where either the orientations alternate or the signs of $\delta Y$ alternate \cite{Su92,Germoni97}.
For example, the representation $5_3 \Succ 7_4 \Succ 5_3' \Prec 3_2 \Prec 5_3''$ is indecomposable.
As observed in the tensor product of two fundamental quartets (section 3.2 and appendix F), this opens the possibility of cycles \cite{Germoni97}
by identification of the end modules of a quiver when they have the same hypercharge.
For example we can identify the modules $5_3$ and $5_3''$ of the previous example
yielding in the notations of \cite{GQS07} the indecomposable cyclic quiver $5_3 \Succ (7_4 \oplus 3_2) \Succ 5_3'$.

As further noticed in \cite{Germoni98},
this opens the possibility of mixing distinct quivers. For example, we can mix
the module  $5_3 \Prec 7_4 \Prec 5_3' \Succ 3_1 \Succ 5_3''$. and a module $3_1 \Succ 5_3'''$ by imposing a
single linear relation between  $5_3$, $5_3''$ and $5_3'''$. It is immediate to verify that these
identifications are allowed, are indecomposable, and are summarized by the off diagonal structure of the $U(1)$ hypercharge $Y$.

As noticed very early, for example in \cite {SNR, Kac2}, in $SU(m/n)$ with
$(m>1,n>1)$ the situation is similar. But since some highest weights can be multiply atypical, such
irreducible modules can be considered as special instances of a multiple parameter indecomposable families
of 2-quivers, one parameter
per atypical direction. But as shown in \cite{Germoni97,Germoni98}, the general classification of the indecomposable modules
  of $SU(m/n)$ is wild.

\subsection {The N generations indecomposable representations.}

Another kind of indecomposable representation was first proposed by Marcu \cite{Marcu80}, mixing two or three copies
of the fundamental typical quartet representation. In 1991 Coquereaux \cite{CQ91} proposed to use these
indecomposable modules of Marcu to give
an algebraic interpretation of the experimental existence of the three families
of elementary particles, associated to the top, the charm and the up quark.
Haussling and Scheck showed later in more details \cite{HS98} how
the free parameters of Marcu correspond to the Pontecorvo, Maki, Nakagawa, Sakata (PMNS) and the 
Cabibbo, Kobayashi and Maskawa (CKM) angles theorized  in the sixties and seventies.
These angles epitomize the misalignment of the strong and weak interactions,
explaining the observed weak decay of the heavy quarks and the matter/antimatter asymmetry of the universe.

Curiously, these representations are not listed in some later integrative analyses of the superalgebra $SU(2/1)$
like the study of G\"{o}tz, Quella, Schomerus \cite{GQS07}.
The reason is that, as explained by Marcu, in these representations the Cartan subalgebra is not diagonalizable,
and such cases are excluded from those later analyses.
Our purpose is to revisit these cases and extend them to any irreducible module
with Dynkin labels $(a,b)$ and any number $N$ of generations.

Consider an irreducible typical module $R(a,b)$ of $SU(2/1)$ of dimension $d$, represented by the matrices $\lX$.
Construct a block diagonal matrix $\LX$ of size $2d$ by repeating $2$ times the matrices $\lX$ along the
diagonal, and add an intertwining matrix $\lX'$ in the top right corner:
\BE
\LX = \begin{pmatrix}
  \lX & \lX'\cr
  0 & \lX
\end{pmatrix}
\EE
Obviously, if $\lX'=0$, the $\LX$ matrices satisfy all the commutation rules of the superalgebra $\ALG$.
Since the $SU(2)$ subalgebra can always be block diagonalized, we look for a solution where
$\lX'$ vanishes for all generators except $Y,U,W$. We have
\BE
H = \begin{pmatrix}
  h & 0\cr
  0 & h
\end{pmatrix}
\;,\;\;
F = \begin{pmatrix}
  f & 0\cr
  0 & f
\end{pmatrix}
\;,\;\;
U = \begin{pmatrix}
  u & u'\cr
  0 & u
\end{pmatrix}
\;,\;\;
V = \begin{pmatrix}
  v & 0\cr
  0 & v
\end{pmatrix}
\;,\;\;
\EE
The conditions $[H,U]=-U$ and $[F,U]=0$ imply that the non-zero entries of $u$ and $u'$ are in the same positions
and that all non-zero entries in any of the four non-zero $u'$  blocks connecting
a pair of even submodules are equal. We now demand that the off diagonal part of the anticommutator $\{U,V\}$
be proportional to the identity
\BE
v u' + u' v = I
\EE
The matrix $v(a,b)$ is known. By inspection, there is a unique solution and the non zero entries of the 4 blocks are respectively $(1,-1,1/(a+1),1/(a+1))$.
(see the example in appendix H). Using $[E,U]=W$, we construct $W$. Using $\{U,V\}=(Y+H)/2$ (2.22), we then find that
\BE
Y = \begin{pmatrix}
  y & 2I\cr
  0 & y
\end{pmatrix}
\EE
and that all the commutation relations of $SU(2/1)$ (2.21-22) are maintained.
Notice that the scale of the off diagonal block can be adjusted by a relative rescaling
of the 2 sectors
\BE
P = \begin{pmatrix} 
   \sqrt{\alpha} & 0\cr
   0 & 1/\sqrt{\alpha}
\end{pmatrix}
\;,\;\;
P^{-1} = \begin{pmatrix} 
   1/\sqrt{\alpha} & 0\cr
   0 & \sqrt{\alpha}
\end{pmatrix}
\;,
\\
\\
PUP^{-1} = \begin{pmatrix} 
   u & \alpha u' \cr
   0 & u
\end{pmatrix}
\;,\;\;
PYP^{-1} = \begin{pmatrix} 
   y & 2 \alpha I \cr
   0 & y
\end{pmatrix}
\;.
\EE

Introducing $N-1$ free complex parameters $(\alpha,\beta,\gamma ...)$, we can immediately generalize the construction to $N$ copies of the
original typical module, where again $(H,E,F,V,X)$ are block diagonal and

\BE
U = \begin{pmatrix}
  u & \alpha u' & 0 & 0 & ...\cr
  0 & u & \beta u' & 0 & ...\cr
  0 & 0 & u & \gamma u' & ...\cr
  ...\cr 
\end{pmatrix}
\;,\;\;\;
Y = \begin{pmatrix}
  y & 2\alpha I & 0 & 0 & ...\cr
  0 & y & 2\beta I & 0 & ...\cr
  0 & 0 & y & 2\gamma I & ...\cr
  ...\cr 
\end{pmatrix}
\EE

According to the classification of Germoni \cite{Germoni98}, all $N$ generations indecomposable representation can be brought to this form.
Our construction is analytic in $b$, so it also covers the limiting atypical indecomposable cases, like the $4^{\Succ}_2=3_2 + 1_0$. 

From a mathematical point of view, the representations with different $\alpha$ are equivalent.
But in particle physics, the $P$ transformation would introduce a relative rescaling
of the free field propagators of the different sectors, which is forbidden. 
This property was used by Haussling, Paschke and Scheck \cite{HPS98} to
interpret the $\alpha$ as the generalized Cabibbo angle.

In a separate study, we prove in a different way the existence of the indecomposable doubling of the Kac modules of $SU(m/n)$ and $OSp(2/2n)$ by studying
the $H^1$ cohomology of the tensor product of the Kac module and its dual \cite{JaTM22a}.
And a little later \cite{TMJG2022b}, we extended the construction (4.9) to the construction of $N-$replications of any Kac module
of any type I Lie-Kac superalgebra.

\subsection {Reduction to the OSp(1/2) sub-superalgebra.}

The embedding (2.23) of $OSp(1/2)$ in $SU(2/1)$ can be applied to the indecomposable
representations. The irreducible atypical representations of $SU(2/1)$ do not split.
The irreducible typical representations of $SU(2/1)$ with their four even submodules
with $SU(2)$ Dynkin labels $(a,a-1,a+1,a)$\,, split into two $OSp(1/2)$ irreducible
representations with $SU(2)$ weights, $(a,a-1)$ and $(a+1,a)$, as if they were atypical.
Hence the indecomposable atypical chains of section 4.1 reduce to a pair of irreducible
$OSp(1/2)$ representations, again as is they were irreducible and typical. The three
$SU(2/1)$ cases: atypical, typical, indecomposable chain, have the same $OSp(1/2)$
decomposition.

The multi-generation indecomposable representations of section 4.2
are more involved. But because each component matrix $(\lX,\mu,\nu)$ splits, the
whole representation can be 
block diagonalized, and the $OSp(1/2)$ representations are fully reducible,
in accordance with \cite{PaisRittenberg75}.

\section{Super-Casimir and chiral super-Casimir operators.} 
\subsection {The classic super-Casimir operators $C_2$ and $C_3$.}

The super-Killing metric is defined as
\BE
g_{AB} = \frac{1}{2} \;STr (\lX_A\lX_B)\;,
\EE
Its non zero entries are
\BE
g_{HH} = -g_{YY} = \alpha\;,\;\;
\\
g_{EF}=g_{FE}=-g_{UV}=g_{VU} =-g_{WX}=g_{XW} = \alpha/2\;,
\\
g^{HH} = -g^{YY} = \alpha^{-1}\;,
\\
g^{EF}=g^{FE}= g^{UV}=-g^{VU} =g^{XW}=-g^{WX} = 2\alpha^{-1}\;.
\EE
To choose the sign of the supertrace, the highest weight is arbitrarily assigned the $(\chi=1, left)$ chirality.
If the representation is typical, $b(b-a-1)\neq 0$, then $\alpha=-(a+1)$.
If $b=0$, atypical 1 representations, then $\alpha = a(a+1)/2$.
If $b=a+1$, atypical 2, then $\alpha = -b(b+1)/2$.

Using the Killing metric $\breve{g}$ of the adjoint representation $(a=1,b=1)$,
the quadratic Casimir operator $C_2$ is defined by the relation
\BE
C_2 = \frac{1}{2} \breve{g}^{AB}\;\lX_A\lX_B = -\frac{1}{4}\;(HH -YY + 2(\{E,F\} + [U,V] + [W,X])\;.
\EE

Although we use the adjoint metric, we define the cubic Casimir operator
$C_3$ in the fundamental representation because it vanishes in the adjoint.
Define
\BE
C_{abc} = \frac{1}{2}\;STr (\lX_a\;\{\lX_b,\lX_c\}),\;
C_{aij} = \frac{1}{2}\;STr (\lX_a\;[\lX_i,\lX_j]),\;
\\
C_{ibj} = \frac{1}{2}\;STr (\lX_i\;\{\lX_a,\lX_j\}),\; 
\;\;\;
C_3 = \frac{1}{6}\;C^{ABC} \;\lX_A\lX_B\lX_C
\;.
\EE
Substituting in this fixed definition the explicit form of the generators $\lX$ computed in section 2 and given in appendices (C,D,E)
we can verify that in any irreducible representation with Dynkin labels $(a,b)$ $C_2$ and $C_3$ are proportional to
the identity with eigenvalues
\BE
 C_2= b(b-a-1) \;I
\\
C_3  = \frac{1}{2} \; b  (b - a -1) (2b - a - 1)\;I
\EE
As expected, these Casimir operators vanish on all atypical representations.
$C_3$ also vanishes if $2b=a+1$, i.e. if the highest hypercharge $Y$ is one: $y=2b-a=1$.
Such representations are `real': they are
equal to their image under the automorphism defined below (2.22).
They include the `real' $OSp(2/2)$ quartet $(a=0,b=1/2)$ (appendix F),
the adjoint $(a=b=1)$ (appendix E) and
the other irreducible weight diagrams symmetric around the $SU(2)$ axis.

Another interesting coefficient is the ratio of the cubic super Casimir tensor $C_{ABC}$ in a generic
representation to the same tensor
$\breve{C}_{ABC}$
computed in the fundamental $(a=1,b=0)$ triplet.
$C_{ABC} = (a+1)(2b -a -1) \breve{C}_{ABC}$.
The scaling for the quark quadruplet
$(a=0,b=2/3)$ is $1/3$, or
$-1/3$ remembering that the quark leading singlet is a right state, and we find that the Casimir tensor
vanishes when we add the contributions of the lepton fundamental triplet (scale 1 by definition) to the contributions of three colored
quark quadruplets. This is known in particle physics as the cancellation of the Adler anomaly by the BIM mechanism (see the references in \cite{TMJ21a} for details).

\subsection {The ghost and the chiral super-Casimir operators $T$ and $T^-$.}

Remember that in a superalgebra, in addition to the identity operator,
there exists a superidentity (or chirality) operator $\chi$\,, which commutes with
all the even generators, anticommutes with all the odd generators
and satisfies $\chi^2 = 1$

Similarly, in addition to the quadratic Casimir operator $C_2$, 
Arnaudon, Bauer and Frappat \cite{ABF97} and Musson {\cite{Musson97} have
discovered in the enveloping algebra ${\rm U}(SU(2/1))$
the existence of a ``ghost'' or chiral super-Casimir operator $T$:
\BE
6\;T = -[U,V] - [W,X]  + \sum \left.(UVWX)\right.
\EE
where the alternated sum ranges over all 24 product of the 4 odd operators.
This operator is interesting in many ways \cite{Gorelik01, DufloPetracci07}.
In \cite{DufloPetracci07} $T$ is called the Gorelik element. In every irreducible representation, one can check by inspection using (2.29-30) that
$T$   is proportional to the quadratic super-Casimir $C_2$ times the superidentity (chirality) operator $\chi$: 
\BE
   T = b (b-a-1)\; \chi = C_2\;\chi \,,
\EE
and vanishes on all atypical representations. This can be checked on a few representations with increasing values of
$(a,b)$ but since we know (appendix E) that $U$ and $W$ are linear in $a$ and $b$ and that $V$ and $X$ are linear in $a$ and
independent of $b$, the identity is valid for all finite dimensional irreducible representations, and later below also
for the indecomposable representations. These eigenvalues were found by Musson {\cite{Musson97} lemma 5.4 for the case of $OSp(1/2n)$.
The identity $\chi^2 = I$ then implies  $T^2 = (C_2)^2\;.$.
  
If we now introduce the operator $T^-$ we can prove by the same method that
\BE
T^- = (UV + WX)(VU + XW) = \frac{1}{2} (1 - \chi) C_2\;.
\EE
In other words, $SU(2/1)$ admits a new  factorizable chiral Casimir operator $T^-$.
This remarkable structure can be generalized to higher superalgebras, but $T^-$ is a Casimir only in $SU(2/1)$. 
 
 If $(a,b,...)$ denote the  even generators of the superalgebra, $(i,j,...)$ the odd generators, and if $(A,B,...)$ and ($I,J,....)$
 denote even and odd monomials in the enveloping algebra, i.e. ordered products of generators containing and even or odd number of odd
 generators, the well known adjoint action $ad$:
 \BE
  ad(a)(A) = [a,A]\;,\;\;ad(a)(J) = [a,J]\;,\\ad(i)(A) = [i,A]\;,\;\;ad(i)(J) = \{i,J\}\;,
 \EE 
 structures the enveloping algebra as a superalgebra representation, and we verify that $ad$ is a representation
 \BE
    [ad(a),ad(b)] = ad([a,b])\;,\;\;[ad(a),ad(i)] = ad([a,i])\;,
 \\
     \{ad(i),ad(j)\} = ad(\{i,j\})\;,\;\;
 \EE
 Arnaudon, Bauer and Frappat \cite{ABF97} have discovered an alternative action $ad'$ of the superalgebra on the enveloping algebra
 which can be defined as
 \BE
 ad'(x)(U) = ad(x)(U\chi)\chi\;\Rightarrow\\
   ad'(a)(A) = [a,A]\;,\;\;ad'(a)(J) = [a,J]\;,\\ad'(i)(A) = \{i,A\}\;,\;\;ad'(i)(J) = [i,J]\;,
 \EE 
   notice that $ad'(i)$ uses the `wrong' type of brackets. We again verify that $ad'$ is a representation
 \BE
   [ad'(a),ad'(b)] = ad'([a,b])\;,\;\;[ad'(a),ad'(i)] = ad'([a,i])\;,\\\{ad'(i),ad'(j)\} = ad'(\{i,j\})\;.\;\;
 \EE
 $ad$ is a derivation obeying the graded Leibniz rule, but $ad'$ obeys a more complicated rule
 \BE
 ad(x)(UV) = ad(x)(U)\;V + (-1)^{|xU|} \;U\;ad(x)(V)\;,\\
 ad'(x)(UV) = ad'(x)(U)\;V - (-1)^{|xU|} \;U\;ad(x)(V)
 \EE
 where $|xU|$ is $1$ if both $x$ and $U$ are odd, otherwise $0$. In the second line, $ad'()$ expands as $ad'() + ad()$, this is not a typo.
 The Casimir operators are invariant under the adjoint action,
 the superidentity $\chi$ and the ghost Casimir (Gorelik element) $T$ are invariant under the alternative action
 \BE
 ad(a)(1) = ad(a)(\chi) = ad'(a)(1) = ad'(a)(\chi) = 0\,\\
 ad(i)(1)=ad(i)(C_2) = 0\;,\;\;ad'(i)(C_2) = 2 i\;C_2\;,\\
 ad(i)(T) = 2 iT\;,\;\;ad'(i)(\chi)=ad'(i)(T) = 0\;.
 \EE

The alternative adjoint action, which acts as a composite of the standard adjoint action and of the chirality operator
seems related to our recent definition of the parallel transport on a super-fiber-bundle. The
standard covariant differential on Lie group manifolds is $D = d + A$, where $d$ is the Cartan differential operator satisfying $d^2=0$, and
$A$ is the connection 1-form valued in the Lie algebra.
The generalization to a superalgebra $D = \chi (d + A) + \Phi$  \cite{TM20a} also requires the presence of the chirality operator
$\chi$ to maintain the desired structure of the curvature 2-form and the Bianchi identity.

\subsection {Casimir operators of the $N$ generations families}

We use the metric $\breve{g}$ of the fundamental representation
and the definition (5.2) of the super-Casimir. Remarkably, the Casimirs becomes block triangular
with all blocks proportional to the $d$-dimensional identity $I_d$, for example if $N=2$ we have:
\BE
\widehat{C_2} =
\begin{pmatrix}
 C_2 \;\;& \alpha C'_2  \cr 0 \;\; & C_2
\end{pmatrix}
\EE
where $C_2 = b (b - a - 1)\; I_d$ and $C'_2 = (2b-a-1)\; I_d$.
Although $\widehat{C_2}$ is not diagonal, it is genuinely a Casimir operator because it
commutes with all the generators of the algebra.
Notice that if we double the indecomposable quartet $4^{\Succ}_2 = 3_3 \Succ 1_0$. Since $b=1,a=0$
the Casimir operator $C_2$ vanishes, as it does on any atypical irreducible module, however $C'_2$
does not vanish.

The chiral super-Casimir operator (5.6) also becomes block triangular
\BE
\widehat{T} =
\begin{pmatrix}
  T \;\;& \alpha T' \cr 0 \;\;& T
\end{pmatrix}
\EE
with each block proportional to the $d$-dimensional
chirality operator $\chi$: we have
$T = b(b-a-1)\chi$ and $T' = (2b-a-1)\chi$
Although $\widehat{T}$ is not diagonal, it is genuinely a ghost Casimir operator because it
commutes with all the even generators of the algebra, and anticommutes with all the odd generators.
$\widehat{C_2}$ and $\widehat{T}$ have the exact same structure and the identity (5.7) is maintained
\BE
 \widehat{T} = \widehat{C_2}\;\chi\;.
\EE

This Cabibbo construction is also valid for the limiting atypical quivers, but then the Casimir operators vanish
along the box diagonal.

In the $N>2$ (generations) indecomposable representations, the identity $T = C_2 \; \chi$ is maintained.
In these cases the second diagonal over the main diagonal is also populated by (scaled) identity or super-identity $\chi$ matrices.
The cubic super-Casimir operator $C_3$  has the same structure with one more non-zero diagonal.
This follows from the recursive nature of the $N$ generations construction given in the previous section
and was checked numerically.
The higher diagonals of the Casimirs $T$, $C_2$ and $C_3$ are computed in appendix H for any values $(a,b,N)$.
By raising the Casimirs to even or odd powers, we can generate higher operators with new linearly independent eigenvalues.
So \textit{in fine}, we can construct a polynomial in these operators globally diagonal and block proportional
to the identity or the superidentity on the whole indecomposable module.

The  (super)center of the enveloping superalgebra is not finitely generated in accordance to
the existence of $N$ generations indecomposable matrices for any $N$.

\section{Resum\'{e}: definitions and main results.}

\noindent
\textbf{Aims and scope:}\\
The aim of this study was to provide explicit constructions of the matrices of the indecomposable representations of the Lie-Kac superalgebra $SU(2/1)$ in the Chevalley basis.
The novelties of our study were that our new proofs are simpler, that we showed the existence of new classes previously thought to be excluded,
and that in each case we gave an actual construction of these representations, something not available in the previous literature.

\noindent
\textbf{Definition:}\\
The complex superalgebra $\ALG = SU(2/1)$ can be defined as the linear span of the eight generators $(F,H,E,U,V,W,X)$ satisfying the
(anti)commutation relations (2.21-22).
A complex finite dimensional representation $\rho$ of $\ALG$ is a linear application $\rho : \ALG \rightarrow End(A)$ of $\ALG$ into the
linear endomorphims of a finite dimensional complex vector space $A$ preserving the commutation relations.
The fundamental representation of $SU(2/1)$ is provided by the $(2/1)$ dimensional matrices with null supertrace (Appendix C). 

\hfill $\Box$

\noindent
\textbf{Theorem 1 : (Kac)}\\
If $A$ is a finite dimensional irreducible representation of $SU(2/1)$, it can be labeled by its Dynkin labels $(a,b)$, where $a$ is a non
negative integer and $b$ is a complex number.

\noindent
\textbf{Structure:}\\
If $a=b=0$, the representation is trivial. The dimension is $1$.\\
If $b=0$, the representation is called atypical 1. It consist of two $SU(2) \oplus U(1)$ modules with Dynkin labels $(a,y=-a)$ and $(a-1,y=-a-1)$ where $y$ is the $U(1)$ eigenvalue of the submodule. The dimension is $(2a+1)$, the superdimension is $1$. \\
If $b=a+1$, the representation is called atypical 2. It splits as $(a,y=a+2)$ and $(a+1,y=a+1)$. The dimension is $(2a+3)$, the superdimension is $-1$.\\
Otherwise, the representation is typical and splits as $(a,y=2b-a)$,  $(a+1,y=2b-a-1)$,  $(a-1,y=2b-a-1)$,  $(a,y=2b-a-2)$. The dimension is $4(a+1)$, the superdimension is $0$. \hfill $\Box$

Notice that the irreducible atypical representation of $SU(2/1)$ are uniquely labeled by their highest $y$ hypercharge, where $y$ is in $\ZZ$.

\noindent
\textbf{Construction:}\\
The matrices $\LX(a,b)$, forming an arbitrary finite dimensional irreducible representation the of $SU(2/1)$ are given explicitly in appendix C

\hfill $\Box$

\noindent
\textbf{Definitions:}\\
If $(A,B,C)$ are complex finite dimensional representations of a Lie superalgebra $\ALG$,
we say that $B$ is an indecomposable 2-quiver denoted $A=B\Prec C$ if there exist an exact sequence
of $\ALG$ modules $0  \rightarrow B \rightarrow A \rightarrow C \rightarrow 0$,  where $C$ is the maximal invariant
submodule of $A$ and $B$ and $C$ are non trivial.
In other words, $A/C = A\; modulo\; C$ is isomorphic to $B$ and $A$ is not isomorphic to $B \oplus C$.
The 2-quiver $B \Prec C$ is called atypical, if the vertex $(B,C)$ are atypical.

\noindent
\textbf{Theorem 2 : (Su, Germoni)}\\
If $(B,C)$ are atypical irreducible representations of $SU(2/1)$ labeled by the $Y$ eigenvalue of their highest weight,
the indecomposable representation $A = B \Prec C$ exists if and only if $y(B) - y(C) = \delta$ where $\delta = \pm 1$. If so, 
the only possible 3 quivers have the form $B_y \Prec C_{y + \delta} \Prec D_y$  or $B_y \Prec C_{y+\delta} \Succ D_{y+2 \delta}$
or $B_y \Succ C_{y+\delta} \Prec D_{y+2\delta}$ where $\delta = \pm 1$.

\hfill $\Box$

\noindent
\textbf{Theorem 3 : (Germoni)}\\
There are no further constraints on atypical $n$-quivers, and they may form multiply-connected oriented quivers (cycles if we drop the orientations of the arcs).
All atypical indecomposable representations of $SU(2/1)$ are of that form.
\noindent
\textbf{Construction:}\\
The matrices representing an arbitrary atypical 2-quiver are given in appendix G. They can be chained to construct $n$-quivers.

\noindent
\textbf{Examples:}\\
The square of the fundamental quartet $4_{2b} \otimes 4_{2b'}$ of $SU(2/1)$ contains if ($b=b'=1/2)$ an indecomposable cycle $(8^{\Succ\oplus\Succ}_2
= 1_0 \Succ (3_2 \oplus 3_{-1}) \Succ 1'_0 $ (Marcu) or if $b=b'=2$ an indecomposable cycle  $12^{\Succ\oplus\Succ}_3=3_2 \Succ (5_3 \oplus 1_0) \Succ 3'_2$
(G\"{o}tz), proving that cycles really occur, in contradiction with theorem 5 of Su.

\hfill $\Box$

\noindent
\textbf{Theorem 4 : $N$ generations indecomposable families (Germoni)}\\ 
If $A$ is an irreducible typical representation, or an atypical quiver, one can construct an indecomposable representation
intertwining $N$ copies of $A$,
where $N$ is any positive integer.

\hfill $\Box$

\noindent
\textbf{Theorem 5 : general doubling (\cite{JaTM22a})}\\
The $N=2$ doubling mechanism can be
extended to all type 1 superalgebras $A(m/n)$ and $C(n)$ 

\noindent
\textbf{Construction: (new)}\\
These matrices are constructed explicitly in section 10 and appendix H.

\hfill $\Box$

\noindent
\textbf{Theorem 6 : Casimirs $T = C_2\;\chi$ (new)}\\
For all finite dimensional irreducible representations, we computed the eigenvalues of the lowest
Casimir operators $C_2$, $C_3$ and $T$ of $SU(2/1)$
and we established the identity $T = C_2\;\chi$,
where $T$ is  the chiral (ghost) Casimir of Arnaudon, also called the Gorelik element,
which is defined (5.6) as a polynomial of mixed degree 2
and 4 in the odd generators. All these Casimirs vanish on the atypical irreducible representations.
The identity was then extended to all the finite dimensional indecomposable representations that we were
able to construct.

\noindent
\textbf{Theorem 7 : Casimirs $T^- = C_2\;(1-\chi)/2$ (new)}\\
For all finite dimensional irreducible representations, we proved that
the quadratic  Casimir operator $C_2$, restricted to the negative chirality states,
i.e. excluding every even layer, is factorizable (5.8),
just in terms of the odd operators $T^- = (UV+WX)(VU+XW)$.

\section{Discussion and conclusions.}

The representation theory of Lie algebras and superalgebras can be
decomposed into three progressively harder problems, the classification,
the characters and the construction of the representations.

For simple Lie algebras, the problem is well understood.
All finite dimensional representations are fully reducible.
The irreducible modules are uniquely labeled by the Dynkin labels of
their highest weight. The characters are given by the Weyl formula.
The matrices can be constructed recursively, at least in principle,
by the repeated action of the simple negative roots on the highest weight.

The classification of the center of the universal covering algebra is
also easy. It is generated by $r$ Casimirs operators, where $r$ is the rank
of the algebra, and each Casimir can be constructed, at least
in principle, by adequate symmetrization of operators of the
form $Tr (\lX_a \lX_b ...) \lX^a \lX^b ...$,  where the indices are raised and lowered
using the invertible Killing metric $g_{ab} = Tr(\lX_a \lX_b)/2$.

The original objective of this study was to clarify the equivalent situation for Lie superalgebras
in view of their application to physics. But the theory is so complex
that we decided to limit our attention to $SU(2/1)$
which is the smallest type 1 basic classical simple superalgebra \cite{Kac0,Kac1}
and has been considered for the classification of the elementary particles \cite{N1,F1}.

As for Lie algebra, its finite dimensional irreducible modules are
classified by the Kac-Dynkin labels $(a,b)$ of their highest weight.
A first difficultly arises in the generalization of the Weyl character formula.
If $b (b-a -1)=0$ the representation is atypical and the Kac module splits
\cite{SNR,Kac0,Kac1,Kac2,TM83,JHKTM90,KacWakimoto94}.
To help with concrete calculations, we established in section 3 and appendix
C, D and E the actual construction of the matrices of all finite dimensional irreducible representations
in the Chevalley basis where all the matrix coefficients can be written in the form $(p + qb)/(a+1)$
where $a$, $p$ and $q$ are integers and $b$ is complex. These quasi-integral matrices are
much easier to manipulate than those presented so far in the literature.

With this tool in hand, we then considered the indecomposable representations.
There is a long history of studies, starting with Marcu \cite{Marcu80} and culminating
with Germoni \cite{Germoni98}. But many of these articles are very abstract and difficult
for a physicist. They sometimes contradict each other on details. All of them are, or at least
seem to be, incomplete. Also, they do provide classifications, but not explicit constructions.

To get a feeling of the problem, we dissected in section (3.2) the decomposition of the product of
two fundamental quartets of $SU(2/1)$ with odd Dynkin labels $b$ and $b'$. 
As already shown in \cite{Marcu80}, this product contains indecomposable cyclic modules.
But depending on the values of $b$ and $b'$, we were led to distinguish more cases than previously considered.

In the classification of finite dimensional indecomposable $SU(2/1)$ modules,
there are two classes.

1) Atypical representations naturally occur as irreducible quotients of
the indecomposable Kac modules, and they naturally form chains because if $\LX$ is an atypical
highest weight, then $\LX + k \beta$ is also atypical, where $\beta$ is the corresponding atypical
root and $k$ an integer. Su \cite{Su92}, completed by Germoni \cite{Germoni97,Germoni98}, has shown that
they can then be combined in oriented graphs, also called quivers, possibly cyclic, following
simple combinatorial rules. Their methods are abstract. We give here a simpler  alternative proof of the
same results, leveraging on our explicit construction of the irreducible matrices.

2) As initially shown by Marcu \cite{Marcu80}, it is also possible to combine two or three
fundamental typical quartets of $SU(2/1)$ into an 8 or a 12 dimensional indecomposable modules.
We generalized, in section 9 and 10,  Marcu's result to the indecomposable mixing of $N$ copies of any irreducible module.
Our proof is very pragmatic, we simply construct the corresponding matrices. As far as we know,
this problem has not much triggered the interest of the mathematicians and we believe
that even the existence of these representations was seldom considered, and that
certainly they had never been constructed explicitly.

Finally, we reviewed in section 5 the discovery by Arnaudon, Bauer and Frappat of the ghost
Casimir operator, which is block-proportional to the superidentity $\chi$ of the superalgebra,
where $\chi$ defines the supertrace as $STr(M) = Tr(\chi M)$. We compute the eigenvalues
of $T$ in the irreducible modules. Then we analyze the non diagonal Casimirs and ghost Casimir operators
in the case of an indecomposable module. Recall that the super-center of the enveloping algebra is
not finitely generated, so we do not list all the Casimir operators but only compute the
lowest ones.
We established the remarkable identity $T = C_2\;\chi$ for all finite dimensional (indecomposable) modules of $SU(2/1)$.
We then showed that the chiral Casimir operator $T^-=C_2(1-\chi)/2)$ is factorizable in term of the odd
generators: $T^- = (UV+WX)(VU+XW)$.

Let us now turn to the implication of this complex zoology for particle physics
\cite {N1,F1,DJ1,NTM1,Jarvis82,NTM82,CQ91,HS98,TM20b,TMJ21a},
$SU(2/1)$ was first considered in 1979 by Ne'eman \cite{N1} and Fairlie \cite{F1}
because the fundamental atypical triplet representation (2.20) matches the quantum numbers
of the electrons and neutrinos, graded by their chirality. Soon after, we found that the fundamental typical quartet
matches the quarks \cite{DJ1,NTM1}. But as noticed by Feynman (1979, private communication),
if the $SU(2/1)$ representations were fully reducible,
the three quark generations would be equivalent, and if
the masses of the quarks are provided by their universal coupling to Higgs fields
associated to the odd generators of the superalgebra, then
the $down$ quark and the $bottom$ quark would have the same mass.
At that time, the $top$ quark was postulated but not yet discovered.
But as noticed by Coquereaux in 1991 \cite{CQ91}, the indecomposable representations
of Marcu \cite{Marcu80} solve this paradox. The intertwining operators
can be interpreted as parameterized by the generalized Cabibbo angles.
Reciprocally, if we interpret in the $SU(2/1)$ framework the difference of mass between the muon and the electron,
via the existence of an indecomposable representation mixing
two generations of leptons, such a mixing can only happen between indecomposable quartets.
In other words, the existence of the more massive muon implies, in $SU(2/1)$, the existence of the right neutrino
\cite{HPS98}.

For a long time, it was thought that one could not extend the construction
of Marcu to the mixing of more than three typical quartet modules,  matching the observed number of generations
of the elementary particles (electron, muon, tau).
This conjecture, mentioned in some of our previous studies \cite{TMJ21a,TM20b},
was too optimistic.
Unfortunately for the physicists,
in  (4.12), we constructed for the first time $N$ generations typical representations
for any positive $N = 2,3,4,5....$.
But on the other hand, we identified a new physically appealing property of $SU(2/1)$.
If we accept the idea that the odd generators of the superalgebra
may be represented in a quantum field theory by scalar fields $\Phi$ \cite{TM20b,TMJ21a},
the structure of the ghost Casimir $T$ (5.6)
exactly matches the standard structure of the Higgs potential, with a quartic interaction $UVWX$,
and a quadratic $UV$ symmetry breaking mass term.

All the equations presented in this work where checked numerically on a large number of examples using a dedicated C-language computer program
 available using the UNIX command 'git clone git@github.com:/ncbi/AceView', then following the instructions in the file AceView/README.su21. 

In conclusion, the representation theory of the simple superalgebra $SU(2/1)$ is very rich,
still incomplete, and interesting for particle physics. We hope that this theory will continue to flourish, for example
in the directions indicated by Duflo and Petracci \cite{DufloPetracci07}.

There remains an open question: 
is this classification truly complete? Could other possibilities exist?
Since so many proposed classifications have been incomplete for so long,
we would not be so surprised if we were still missing alternative constructions of a different kind.

\section*{Acknowledgments.}
This research was supported by the Intramural Research Program of the National Library of Medicine, National Institute of Health.
We are grateful to Danielle Thierry-Mieg for clarifying the presentation.

\begin{appendix}
\section{Finite dimensional irreducible representations of $SU(2)$.}

Applying the results of section 2.1,
the 4 dimensional representation of $SU(2)$ is given by the matrices
\BE
H = \begin{pmatrix} 
 3 & 0 & 0 & 0 \cr 0 & 1 & 0 & 0 \cr 0 & 0 & -1 & 0 
 \cr 0 & 0 & 0 & -3 
\end{pmatrix}
\;,\;\;\;
F = \begin{pmatrix} 
 0 & 0 & 0 & 0 \cr 1 & 0 & 0 & 0 \cr 0 & 1 & 0 & 0 
 \cr 0 & 0 & 1 & 0
\end{pmatrix}
\;,\;\;\;
E = \begin{pmatrix} 
 0 & 3 & 0 & 0 \cr 0 & 0 & 4 & 0 \cr 0 & 0 & 0 & 3 
 \cr 0 & 0 & 0 & 0
\end{pmatrix}
\;,\;\;\;
\EE
All finite dimensional dimensional irreducible representations of $SU(2)$ are of a similar form,
with coefficients given by equations (2.4) and (2.5). 

\section{Finite dimensional irreducible representations of $OSp(1/2)$.}

Applying the results of section 2.2,
we show the matrices for the finite dimensional
irreducible representation with Dynkin label $a=3$:
\BE
H = \begin{pmatrix} 
 3 & 0 & 0 & 0 & 0 & 0 & 0  \cr 0 & 1 & 0 & 0 & 0 & 0 & 0 \cr 0 & 0 & -1 & 0 & 0 & 0 & 0  
 \cr 0 & 0 & 0 & -3 & 0 & 0 & 0  
\cr 0 & 0 & 0 & 0 & 2 & 0 & 0
 \cr 0 & 0 & 0 & 0 & 0 & 0 & 0
 \cr 0 & 0 & 0 & 0 & 0 & 0 & -2
 \end{pmatrix}
\;,\;\;\;
F = \begin{pmatrix} 
 0 & 0 & 0 & 0 & 0 & 0 & 0 \cr 1 & 0 & 0 & 0 & 0 & 0 & 0 \cr 0 & 1 & 0 & 0 & 0 & 0 & 0 
 \cr 0 & 0 & 1 & 0 & 0 & 0 & 0 
\cr 0 & 0 & 0 & 0 & 0 & 0 & 0
 \cr 0 & 0 & 0 & 0 & 1 & 0 & 0
 \cr 0 & 0 & 0 & 0 & 0 & 1 & 0
 \end{pmatrix}
\;,\;\;\;
E = \begin{pmatrix} 
 0 & 3 & 0 & 0 & 0 & 0 & 0 \cr 0 & 0 & 4 & 0 & 0 & 0 & 0 \cr 0 & 0 & 0 & 3 & 0 & 0 & 0 
 \cr 0 & 0 & 0 & 0 & 0 & 0 & 0 
\cr 0 & 0 & 0 & 0 & 0 & 2 & 0
 \cr 0 & 0 & 0 & 0 & 0 & 0 & 2
 \cr 0 & 0 & 0 & 0 & 0 & 0 & 0
\end{pmatrix}
\;,\;\;\;
\\
\\
\\
R = \begin{pmatrix} 
 0 & 0 & 0 & 0 & 3 & 0 & 0 \cr 0 & 0 & 0 & 0 & 0 & 2 & 0 \cr 0 & 0 & 0 & 0 & 0 & 0 & 1
 \cr 0 & 0 & 0 & 0 & 0 & 0 & 0
 \cr 0 & 1 & 0 & 0 & 0 & 0 & 0
 \cr 0 & 0 & 2 & 0 & 0 & 0 & 0
 \cr 0 & 0 & 0 & 3 & 0 & 0 & 0
 \end{pmatrix}
\;,\;\;\;
S = \begin{pmatrix} 
 0 & 0 & 0 & 0 & 0 & 0 & 0 \cr 0 & 0 & 0 & 0 & 1 & 0 & 0 \cr 0 & 0 & 0 & 0 & 0 & 1 & 0
 \cr 0 & 0 & 0 & 0 & 0 & 0 & 1
 \cr -1 & 0 & 0 & 0 & 0 & 0 & 0
 \cr 0 & -1 & 0 & 0 & 0 & 0 & 0
 \cr 0 & 0 & -1 & 0 & 0 & 0 & 0
 
\end{pmatrix}
\;,\;\;\;
\EE
All finite dimensional irreducible representations of $OSp(1/2)$ are of this form.
Just extend the diagonals of $S$ with more $1$ or $-1$ and the diagonals of $R$ with
numbers up to $a$.

\section{Atypical 1 irreducible representations of $SU(2/1)$.}

As an example. the irreducible finite dimensional atypical 1 representation of $SU(2/1)$
of highest weight $|3,0>$ is given by the matrices
\BE
H = \begin{pmatrix} 
 3 & 0 & 0 & 0 & 0 & 0 & 0  \cr 0 & 1 & 0 & 0 & 0 & 0 & 0 \cr 0 & 0 & -1 & 0 & 0 & 0 & 0  
 \cr 0 & 0 & 0 & -3 & 0 & 0 & 0  
\cr 0 & 0 & 0 & 0 & 2 & 0 & 0
 \cr 0 & 0 & 0 & 0 & 0 & 0 & 0
 \cr 0 & 0 & 0 & 0 & 0 & 0 & -2
 \end{pmatrix}
\;,\;\;
Y = \begin{pmatrix} 
 -3 & 0 & 0 & 0 & 0 & 0 & 0 \cr 0 & -3 & 0 & 0 & 0 & 0 & 0 \cr 0 & 0 & -3 & 0 & 0 & 0 & 0 
 \cr 0 & 0 & 0 & -3 & 0 & 0 & 0 
\cr 0 & 0 & 0 & 0 & -4 & 0 & 0
 \cr 0 & 0 & 0 & 0 & 0 & -4 & 0
 \cr 0 & 0 & 0 & 0 & 0 & 0 & -4
 \end{pmatrix}
\;,\;\;
\\
\\
\\
F = \begin{pmatrix} 
 0 & 0 & 0 & 0 & 0 & 0 & 0 \cr 1 & 0 & 0 & 0 & 0 & 0 & 0 \cr 0 & 1 & 0 & 0 & 0 & 0 & 0 
 \cr 0 & 0 & 1 & 0 & 0 & 0 & 0 
\cr 0 & 0 & 0 & 0 & 0 & 0 & 0
 \cr 0 & 0 & 0 & 0 & 1 & 0 & 0
 \cr 0 & 0 & 0 & 0 & 0 & 1 & 0
 \end{pmatrix}
\;,\;\;
E = \begin{pmatrix} 
 0 & 3 & 0 & 0 & 0 & 0 & 0 \cr 0 & 0 & 4 & 0 & 0 & 0 & 0 \cr 0 & 0 & 0 & 3 & 0 & 0 & 0 
 \cr 0 & 0 & 0 & 0 & 0 & 0 & 0 
\cr 0 & 0 & 0 & 0 & 0 & 2 & 0
 \cr 0 & 0 & 0 & 0 & 0 & 0 & 2
 \cr 0 & 0 & 0 & 0 & 0 & 0 & 0
\end{pmatrix}
\;,\;\;\;
\\
\\
\\
U = \begin{pmatrix} 
 0 & 0 & 0 & 0 & 0 & 0 & 0 \cr 0 & 0 & 0 & 0 & -1 & 0 & 0 \cr 0 & 0 & 0 & 0 & 0 & -1 & 0
 \cr 0 & 0 & 0 & 0 & 0 & 0 & -1
 \cr 0 & 0 & 0 & 0 & 0 & 0 & 0
 \cr 0 & 0 & 0 & 0 & 0 & 0 & 0
 \cr 0 & 0 & 0 & 0 & 0 & 0 & 0
 \end{pmatrix}
\;,\;\;
V = \begin{pmatrix} 
 0 & 0 & 0 & 0 & 0 & 0 & 0 \cr 0 & 0 & 0 & 0 & 0 & 0 & 0 \cr 0 & 0 & 0 & 0 & 0 & 0 & 0
 \cr 0 & 0 & 0 & 0 & 0 & 0 & 0
 \cr 0 & 1 & 0 & 0 & 0 & 0 & 0
 \cr 0 & 0 & 2 & 0 & 0 & 0 & 0
 \cr 0 & 0 & 0 & 3 & 0 & 0 & 0
 
\end{pmatrix}
\;,\;\;
\\
\EE
\BE
\\
W = \begin{pmatrix} 
 0 & 0 & 0 & 0 & -3 & 0 & 0 \cr 0 & 0 & 0 & 0 & 0 & -2 & 0 \cr 0 & 0 & 0 & 0 & 0 & 0 & -1
 \cr 0 & 0 & 0 & 0 & 0 & 0 & 0
 \cr 0 & 0 & 0 & 0 & 0 & 0 & 0
 \cr 0 & 0 & 0 & 0 & 0 & 0 & 0
 \cr 0 & 0 & 0 & 0 & 0 & 0 & 0
 \end{pmatrix}
\;,\;\;
X = \begin{pmatrix} 
 0 & 0 & 0 & 0 & 0 & 0 & 0 \cr 0 & 0 & 0 & 0 & 0 & 0 & 0 \cr 0 & 0 & 0 & 0 & 0 & 0 & 0
 \cr 0 & 0 & 0 & 0 & 0 & 0 & 0
 \cr 1 & 0 & 0 & 0 & 0 & 0 & 0
 \cr 0 & 1 & 0 & 0 & 0 & 0 & 0
 \cr 0 & 0 & 1 & 0 & 0 & 0 & 0
 
\end{pmatrix}
\;,\;\;\;
\EE
\noindent
the general irreducible atypical 1 representation, with highest weight $|a,0>$ ($a$ integer), has the same shape extending the
enumerations $(1,2,3)$ to $(1,2,...a)$ in $(V,W)$ and using the $SU(2)$
matrices $E_{p,p+1} = p(a'-p+1),\;p=1,...a'$ in the two diagonal blocks $a'=a$ and $a'=a-1$.

\section{Atypical 2 irreducible representations of $SU(2/1)$.}

The general atypical 2 matrices are very similar.
The irreducible finite dimensional atypical 2 representation of $SU(2/1)$
of highest weight $|2,3>$ is given by the matrices
\BE
H = \begin{pmatrix} 
 2 & 0 & 0 & 0 & 0 & 0 & 0  \cr 0 & 0 & 0 & 0 & 0 & 0 & 0 \cr 0 & 0 & -2 & 0 & 0 & 0 & 0  
 \cr 0 & 0 & 0 & 3 & 0 & 0 & 0  
\cr 0 & 0 & 0 & 0 & 1 & 0 & 0
 \cr 0 & 0 & 0 & 0 & 0 & -1 & 0
 \cr 0 & 0 & 0 & 0 & 0 & 0 & -3
 \end{pmatrix}
\;,\;\;
Y = \begin{pmatrix} 
 4 & 0 & 0 & 0 & 0 & 0 & 0 \cr 0 & 4 & 0 & 0 & 0 & 0 & 0 \cr 0 & 0 & 4 & 0 & 0 & 0 & 0 
 \cr 0 & 0 & 0 & 3 & 0 & 0 & 0 
\cr 0 & 0 & 0 & 0 & 3 & 0 & 0
 \cr 0 & 0 & 0 & 0 & 0 & 3 & 0
 \cr 0 & 0 & 0 & 0 & 0 & 0 & 3
 \end{pmatrix}
\;,\;\;
\\
\\
\\
F = \begin{pmatrix} 
 0 & 0 & 0 & 0 & 0 & 0 & 0 \cr 1 & 0 & 0 & 0 & 0 & 0 & 0 \cr 0 & 1 & 0 & 0 & 0 & 0 & 0 
 \cr 0 & 0 & 0 & 0 & 0 & 0 & 0 
\cr 0 & 0 & 0 & 1 & 0 & 0 & 0
 \cr 0 & 0 & 0 & 0 & 1 & 0 & 0
 \cr 0 & 0 & 0 & 0 & 0 & 1 & 0
 \end{pmatrix}
\;,\;\;
E = \begin{pmatrix} 
 0 & 2 & 0 & 0 & 0 & 0 & 0 \cr 0 & 0 & 2 & 0 & 0 & 0 & 0 \cr 0 & 0 & 0 & 0 & 0 & 0 & 0 
 \cr 0 & 0 & 0 & 0 & 3 & 0 & 0 
\cr 0 & 0 & 0 & 0 & 0 & 4 & 0
 \cr 0 & 0 & 0 & 0 & 0 & 0 & 3
 \cr 0 & 0 & 0 & 0 & 0 & 0 & 0
\end{pmatrix}
\;,\;\;\;
\\
\\
\\
U = \begin{pmatrix} 
 0 & 0 & 0 & 1 & 0 & 0 & 0 \cr 0 & 0 & 0 & 0 & 1 & 0 & 0 \cr 0 & 0 & 0 & 0 & 0 & 1 & 0
 \cr 0 & 0 & 0 & 0 & 0 & 0 & 0
 \cr 0 & 0 & 0 & 0 & 0 & 0 & 0
 \cr 0 & 0 & 0 & 0 & 0 & 0 & 0
 \cr 0 & 0 & 0 & 0 & 0 & 0 & 0
 \end{pmatrix}
\;,\;\;
V = \begin{pmatrix} 
 0 & 0 & 0 & 0 & 0 & 0 & 0 \cr 0 & 0 & 0 & 0 & 0 & 0 & 0 \cr 0 & 0 & 0 & 0 & 0 & 0 & 0
 \cr 3 & 0 & 0 & 0 & 0 & 0 & 0
 \cr 0 & 2 & 0 & 0 & 0 & 0 & 0
 \cr 0 & 0 & 1 & 0 & 0 & 0 & 0
 \cr 0 & 0 & 0 & 0 & 0 & 0 & 0
 
\end{pmatrix}
\;,\;\;\;
\\
\\
\\
W = \begin{pmatrix} 
 0 & 0 & 0 & 0 & -1 & 0 & 0 \cr 0 & 0 & 0 & 0 & 0 & -2 & 0 \cr 0 & 0 & 0 & 0 & 0 & 0 & -3
 \cr 0 & 0 & 0 & 0 & 0 & 0 & 0
 \cr 0 & 0 & 0 & 0 & 0 & 0 & 0
 \cr 0 & 0 & 0 & 0 & 0 & 0 & 0
 \cr 0 & 0 & 0 & 0 & 0 & 0 & 0
 \end{pmatrix}
\;,\;\;
X = \begin{pmatrix} 
 0 & 0 & 0 & 0 & 0 & 0 & 0 \cr 0 & 0 & 0 & 0 & 0 & 0 & 0 \cr 0 & 0 & 0 & 0 & 0 & 0 & 0
 \cr 0 & 0 & 0 & 0 & 0 & 0 & 0
 \cr -1 & 0 & 0 & 0 & 0 & 0 & 0
 \cr 0 & -1 & 0 & 0 & 0 & 0 & 0
 \cr 0 & 0 & -1 & 0 & 0 & 0 & 0
 
\end{pmatrix}
\;,\;\;
\EE

the general irreducible atypical 2 representation, with highest weight $|a,a+1>$ ($a$ integer), has the same shape extending the
enumerations $(1,2,3)$ to $(1,2,...a+1)$ in $(V,W)$ and using the $SU(2)$
matrices $E_{p,p+1} = p(a'-p+1),\;p=1,...a'$ in the two diagonal blocks $a'=a$ and $a'= b = a+1$.

\section{Example: the shifted adjoint typical representation.} 

Notice that in the typical case $b(b-a-1)\neq 0)$, the Kac-Dynkin label $(b)$ is not quantized.
For example, we can start from the adjoint representation with highest weight $|1,1>$, shift $(b)$
and obtain another 8 dimensional representation given by the matrices \\
\mbox{}\hskip-.5cm
\BE
\hskip-.5cm H = \begin{pmatrix} 
  1 & 0 & 0 & 0 & 0 & 0 & 0
 \cr 0 & -1 & 0 & 0 & 0 & 0 & 0 & 0

 \cr 0 & 0 & 2 & 0 & 0 & 0 & 0 & 0  
 \cr 0 & 0 & 0 & 0 & 0 & 0 & 0 & 0
  \cr 0 & 0 & 0 & 0 & -2 & 0 & 0 & 0
 
  \cr 0 & 0 & 0 & 0 & 0 & 0 & 0 & 0
  
 \cr 0 & 0 & 0 & 0 & 0 & 0 & 1 & 0
 \cr 0 & 0 & 0 & 0 & 0 & 0 & 0 & -1
 \end{pmatrix}
\;,\;\;\\ \\ \\
Y = \begin{pmatrix} 
  2b\!-\!1 \!\!&\!\! 0 \!\!&\!\! 0 \!\!&\!\! 0 \!\!&\!\! 0 \!\!&\!\! 0 \!\!&\!\! 0 \!\!&\!\! 0
 \cr 0 & 2b\!-\!1 & 0 & 0 & 0 & 0 & 0 & 0

 \cr 0 & 0 & 2b\!-\!2 & 0 & 0 & 0 & 0 & 0  
 \cr 0 & 0 & 0 & 2b\!-\!2 & 0 & 0 & 0 & 0
 \cr 0 & 0 & 0 & 0 & 2b\!-\!2& 0 & 0 & 0
 
 \cr 0 & 0 & 0 & 0 & 0 & 2b\!-\!2& 0 & 0
  
 \cr 0 & 0 & 0 & 0 & 0 & 0 &  2b\!-\!3& 0
 \cr 0 & 0 & 0 & 0 & 0 & 0 & 0 & 2b\!-\!3
 \end{pmatrix}
\;,\;\;
\\
\\
\\
F = \begin{pmatrix} 
  0 & 0 & 0 & 0 & 0 & 0 & 0
 \cr 1 & 0 & 0 & 0 & 0 & 0 & 0 & 0

 \cr 0 & 0 & 0 & 0 & 0 & 0 & 0 & 0  
 \cr 0 & 0 & 1 & 0 & 0 & 0 & 0 & 0
  \cr 0 & 0 & 0 & 1 & 0 & 0 & 0 & 0
 
  \cr 0 & 0 & 0 & 0 & 0 & 0 & 0 & 0
  
 \cr 0 & 0 & 0 & 0 & 0 & 0 & 0 & 0
 \cr 0 & 0 & 0 & 0 & 0 & 0 & 1 & 0
 \end{pmatrix}
\;,\;\;
E = \begin{pmatrix} 
  0 & 1 & 0 & 0 & 0 & 0 & 0
 \cr 0 & 0 & 0 & 0 & 0 & 0 & 0 & 0

 \cr 0 & 0 & 0 & 2 & 0 & 0 & 0 & 0  
 \cr 0 & 0 & 0 & 0 & 2 & 0 & 0 & 0
  \cr 0 & 0 & 0 & 0 & 0 & 0 & 0 & 0
 
  \cr 0 & 0 & 0 & 0 & 0 & 0 & 0 & 0
  
 \cr 0 & 0 & 0 & 0 & 0 & 0 & 0 & 1
 \cr 0 & 0 & 0 & 0 & 0 & 0 & 0 & 0
\end{pmatrix}
\;,\;\;\;
\\
\\
\\
\EE

\BE
U(1,b) = \begin{pmatrix} 
  0 & 0 & b & 0 & 0 & 0 & 0 & 0
  \cr 0 & 0 & 0 & b & 0 & 2-b & 0 & 0

  \cr 0 & 0 & 0 & 0 & 0 & 0 & 0 & 0
  \cr 0 & 0 & 0 & 0 & 0 & 0 & b/2-1 & 0
  \cr 0 & 0 & 0 & 0 & 0 & 0 & 0 & b/2-1
  
  \cr 0 & 0 & 0 & 0 & 0 & 0 & b/2 & 0
  
 \cr 0 & 0 & 0 & 0 & 0 & 0 & 0 & 0
 \cr 0 & 0 & 0 & 0 & 0 & 0 & 0 & 0
 \end{pmatrix}
\;,\;\;
V(1,b) = \begin{pmatrix} 
      0 & 0 & 0 & 0 & 0 & 0 & 0 & 0
  \cr 0 & 0 & 0 & 0 & 0 & 0 & 0 & 0
  
  \cr 1 & 0 & 0 & 0 & 0 & 0 & 0 & 0
  \cr 0 & 1/2 & 0 & 0 & 0 & 0 & 0 & 0
  \cr 0 & 0 & 0 & 0 & 0 & 0 & 0 & 0

  \cr 0 & -1/2 & 0 & 0 & 0 & 0 & 0 & 0
  
  \cr 0 & 0 & 0 & 1 & 0 & 1 & 0 &  0
  \cr 0 & 0 & 0 & 0 & 2 & 0 & 0  & 0 
 
\end{pmatrix}
\;,\;\;\;
\\
\\
\\
\;,\;\;
\EE

As another example, we give the matrices $(U,V)$ for the representation with highest weight $|2,b>$
using the notations $c=b/3$ and $d=1/3$
\BE
U(2,b) = \begin{pmatrix} 
      0\;0\;0 & b\;0\;0\;0 & 0   & 0   & 0 & 0 & 0 
  \cr 0\;0\;0 & 0\;b\;0\;0 & 3-b & 0   & 0 & 0 & 0
  \cr 0\;0\;0 & 0\;0\;b\;0 & 0   & 3-b & 0 & 0 & 0

  \cr 0\;0\;0\; & 0\;0\;0\;0 & 0 & 0 & 0 & 0 & 0
  \cr 0\;0\;0\; & 0\;0\;0\;0 & 0 & 0 & c-1 & 0 & 0
  \cr 0\;0\;0\; & 0\;0\;0\;0 & 0 & 0 & 0 & c-1 & 0
  \cr 0\;0\;0\; & 0\;0\;0\;0 & 0 & 0 & 0 & 0 & c-1

  \cr 0\;0\;0\; & 0\;0\;0\;0 & 0 & 0 & c & 0 & 0
  \cr 0\;0\;0\; & 0\;0\;0\;0 & 0 & 0 & 0 & c & 0

  \cr 0\;0\;0\; & 0\;0\;0\;0 & 0 & 0 & 0 & 0 & 0
  \cr 0\;0\;0\; & 0\;0\;0\;0 & 0 & 0 & 0 & 0 & 0
  \cr 0\;0\;0\; & 0\;0\;0\;0 & 0 & 0 & 0 & 0 & 0
 \end{pmatrix}
\;,\;\;
\EE

\BE
V(2,b) = \begin{pmatrix} 
      0 & 0 & 0 & 0 & 0\; 0\; 0\; 0 & 0 & 0\; 0\; 0
  \cr 0 & 0 & 0 & 0 & 0\; 0\; 0\; 0 & 0 & 0\; 0\; 0
  \cr 0 & 0 & 0 & 0 & 0\; 0\; 0\; 0 & 0 & 0\; 0\; 0

  \cr 3d & 0 & 0 & 0 & 0\; 0\; 0\; 0 & 0& 0\; 0\; 0
  \cr 0 & 2d & 0 & 0 & 0\; 0\; 0\; 0 & 0& 0\; 0\; 0
  \cr 0 & 0  & d & 0 & 0\; 0\; 0\; 0 & 0& 0\; 0\; 0
  \cr 0 & 0  & 0 & 0 & 0\; 0\; 0\; 0 & 0& 0\; 0\; 0
  
  \cr 0 & -d & 0 & 0 & 0\; 0\; 0\; 0 & 0& 0\; 0\; 0
  \cr 0 & 0 & -2d &0 & 0\; 0\; 0\; 0 & 0& 0\; 0\; 0
  
  \cr 0 & 0 &  0 & 0 & 1\; 0\; 0\; 2 & 0& 0\; 0\; 0
  \cr 0 & 0 & 0 & 0 &  0\; 2\; 0\; 0 & 1& 0\; 0\; 0
  \cr 0 & 0 & 0 & 0 &  0\; 0\; 3 \; 0  &0& 0\; 0\; 0
\end{pmatrix}
\;,\;\;\;
\\
\\
\\
\;,\;\;
\EE

\section{The tensor product of two quartets is indecomposable.}

Consider a quartet representation of $SU(2/1)$ with Dynkin labels $(0,b)$::

\BE
\hskip-.5cm
H = \begin{pmatrix} 
  0 & 0 & 0 & 0 \cr
  0 & 1 & 0 & 0 \cr
  0 & 0 & -1 & 0 \cr
  0 & 0 & 0 & 0 \cr
 \end{pmatrix}
\;,\;\;
Y = \begin{pmatrix} 
  2b & 0 & 0 & 0 \cr
  0 & 2b-1 & 0 & 0 \cr
  0 & 0 & 2b-1 & 0 \cr
  0 & 0 & 0 & 2b-2 \cr
 \end{pmatrix}
\;,\;\;
\\
\\
\\
\hskip-.5cm F = \begin{pmatrix} 
  0 & 0 & 0 & 0 \cr
  0 & 0 & 0 & 0 \cr
  0 & 1 & 0 & 0 \cr
  0 & 0 & 0 & 0 \cr
 \end{pmatrix}
\;,\;\;
E = \begin{pmatrix} 
  0 & 0 & 0 & 0 \cr
  0 & 0 & 1 & 0 \cr
  0 & 0 & 0 & 0 \cr
  0 & 0 & 0 & 0 \cr
 \end{pmatrix}
\;,\;\;
\\
\\
\hskip-1cm U = \begin{pmatrix} 
  0 & b & 0 & 0 \cr
  0 & 0 & 0 & 0 \cr
  0 & 0 & 0 & b-1 \cr
  0 & 0 & 0 & 0 \cr
 \end{pmatrix}
\;,\;\;
V = \begin{pmatrix} 
  0 & 0 & 0 & 0 \cr
  1 & 0 & 0 & 0 \cr
  0 & 0 & 0 & 0 \cr
  0 & 0 & 1 & 0 \cr
 \end{pmatrix}
        \\ \\
W = \begin{pmatrix} 
  0 & 0 & -b & 0 \cr
  0 & 0 & 0 & b-1 \cr
  0 & 0 & 0 & 0 \cr
  0 & 0 & 0 & 0 \cr
 \end{pmatrix}
\;,\;\;
X = \begin{pmatrix} 
  0 & 0 & 0 & 0 \cr
  0 & 0 & 0 & 0 \cr
  -1 & 0 & 0 & 0 \cr
  0 & 1 & 0 & 0 \cr
 \end{pmatrix}
\;,\;\;
\EE

Generically, this representation is irreducible and we denote it $4_{2b}$
where the lower index denotes the
highest eigenvalue of the $Y$ operator.
But if $b=0$, atypical $1$, the top singlet generates the triplet, but is not in the orbit of the triplet.
This representation is indecomposable and we denote it $4^{\Succ}_0 = 1_0 \Succ 3_{-1}$.
If $b=2$, atypical $2$, the top triplet generates the bottom singlet and which acts as a mousetrap.
We denote this case $4^{\Succ}_2=3_2 \Succ 1_0$.
The reciprocal cases  $4^{\Prec}_0 = 1_0 \Prec 3_{-1}$, respectively $4^{\Prec}_2=3_2 \Prec 1_0$,  are not covered in the above notation. They exist
as lowest weight modules for which some components of
$V$ vanish and are given by
\BE
U = \begin{pmatrix} 
  0 & 1 & 0 & 0 \cr
  0 & 0 & 0 & 0 \cr
  0 & 0 & 0 & 1 \cr
  0 & 0 & 0 & 0 \cr
 \end{pmatrix}
\;,\;\;
V = \begin{pmatrix} 
  0 & 0 & 0 & 0 \cr
  c & 0 & 0 & 0 \cr
  0 & 0 & 0 & 0 \cr
  0 & 0 & 1-c & 0 \cr
 \end{pmatrix}
\;,\;\;
\EE
with $c=0$, respectively $c=1$.

Consider now the tensor product of two quartet representations $4^{\Succ}_{2b}\;\otimes\;4^{\Succ}_{2b'}$ with Dynkin labels $(0,b)$ and $(0,b')$.
Call $(k,\lX,\mu,n)$ the states of the quartet, given the matrices (F.1), we can write
\BE
  |\lX> = V|k>\;,\;\;|\mu>=FV|k>\;,\;\;|n>=FVF|k>\;. 
\EE
Generically, the tensor product contains 16 states organized as a $4_{2(b+b')}$,  a $4_{2(b+b'-1)}$ and a shifted
adjoint $8_{2(b+b')-1}$ 
but the devil hides in the details, if $(b+b')= 0\;or\;1\;or\;2$, the quartets are reducible but indecomposable and if
$(b+b') = 2 \;or\;0$ the octet is reducible but indecomposable and we must distinguish the cases
$(3 \Prec 1)$,
$(3 \Succ 1)$,
$(3 \Prec 5)$,
$(3 \Succ 5)$. 
To distinguish these cases, we shall compute the orbit of the extreme states $|kk>$, $|nn>$, $|\lX\lX>$ and $|\mu\mu>$.
$|kk>$ and $|nn>$ are $SU(2)$ singlets and the 3 states $|\lX\lX>$, $|\lX\mu+\mu\lX>$ and $2|\mu\mu>$ form
and $SU(2)$ triplet. 

The orbit of the highest weight $|kk>$ of the tensor product is generated by the four states
\BE
|kk>\;,\,\,V|kk>=|\lX k + k \lX>\;,\;\;FV|kk>=|\mu k+k\mu>\;,
\\
|\omega>=VFV|kk>=|nk+kn-\mu\lX+\lX\mu>\;,\;\;F|\omega>=V|\omega>=0\;.
\EE
where we used the fact that $V$ acting on $|\mu k>$ generates a minus sign when traversing the odd state $\mu$.

We now compute the action of the raising operators and verify that
\BE
E |kk> = EV|kk>=E|\omega>=0\;,\;\;EFV|kk>=V|kk>\;,
\\
U|kk>=UFV|kk>=0\;,\;\;UV|kk>=(b+b')|kk>\;,\
\\
U|\omega>=(b+b'-1)|\mu k + k\mu> = (b+b'-1) FV|kk>\;.
\EE
The orbit of $|kk>$ is the expected quartet with Dynkin labels $(0,b+b')$. If $b+b'=0$ we have an indecomposable
$4^{\Succ}_0 = 1_0 \Succ 3_{-1}$
if $b+b'=1$ we have an indecomposable $4^{\Succ}_2= 3_2 \Succ 1_0$, otherwise we have an irreducible quartet $4_{2(b+b')}$.
We now consider the orbit of the lowest weight $|n n>$ which is generated by the 4 states:
\BE
|nn>\;,\,\,U|nn>=(b-1)|\mu n> + (b'-1)|n\mu>\;,\;\;\\\,EU|nn>=(b-1)|\lX n>+(b'-1)|n\lX>\;,
\\
|\overline{\omega}>=UEU|nn>\\=b(b-1)|kn>+b'(b'-1)|nk>+(b-1)(b'-1)|\mu\lX - \lX\mu>\;,
\\
F|\overline{\omega}> = U|\overline{\omega}> = 0\;.

\EE
Then we check if the representation is irreducible or indecomposable:
\BE
F |nn> = FU|nn>=F|\overline{\omega}>=0\;,\;\;FEU|nn>=U|nn>\;,
\\
V|nn>=VEU|nn>=0\;,\;\;VU|nn>=(b+b'-2)|nn>\;,\
\\
V|\overline{\omega}>=b(b-1)|\lX n>+b'(b'-1)|n\lX>+(b-1)(b'-1)|n\lX+\lX n>\;
\\
= (b-1)(b+b'-1)|\lX n> + (b'-1)(b+b'-1)|n\lX>
\\
= (b+b'-1)EU|nn>\;.
\EE
We have a lowest weight quartet. If $b+b'=2$, the representation is an indecomposable $4^{\Prec}_2= 3_2 \Prec 1_0$. If $b+b'=1$
we have an indecomposable $4^{\Prec}_0= 1_0 \Prec 3_{-1}$, otherwise we have an irreducible quartet.

We then compare the $SU(2)$ singlets  $\omega$ and $\overline{\omega}$. If $(b+b'=1)$ both are at the same time
highest weights and lowest weights and they actually coincide.
\BE
b+b'=1 \\
\Rightarrow \overline{\omega} = b(b-1)|k n>+(1-b)(-b)|nk>+(b-1)(-b)|\mu\lX - \lX\mu>\;\\
= b(b-1) |kn + nk + \lX\mu - \mu\lX> \\
= b(b-1)\omega\;.
\EE
Therefore, if $b+b'=1$ and $b(b-1) \neq 0$, we have constructed only 7 states and they generate a single indecomposable $7_2^{\Succ\Prec} = 3_2\Succ 1_0 \Prec 3_{-1}$.
But this also implies that we are still missing one $SU(2)$ symmetric state, namely
\BE
\breve\omega = |kn + nk - \lX\mu + \mu\lX>\;,\;\;F \breve\omega = H  \breve\omega = E \breve\omega = 0.
\EE
And as shown in \cite{GQS07}, using $b+b'=1$, this singlet generates the two triplets as one way traps
\BE
V\breve\omega = 2|\lX n + n\lX>\;,\\
U\breve\omega = 2(y|k \mu +\mu k>\;,\\
UV\breve\omega = VU\breve\omega = 0
\EE
Thus our $7$ is actually an invariant submodule of a cyclic indecomposable representation of dimension $8$ which can be presented as
$8_2^{\Succ \oplus \Succ} = 1_0 \Succ (3_2 \oplus 3_{-1})  \Succ 1'_0$,
where the $1_0$ generated by $\breve\omega$) is distinct from $1'_0$  generated by $\omega=\overline\omega$).
Notice however that if $(b=1 \Rightarrow b'=0)$ the triplet starting on $|nn>$ is irreducible
So we have an acyclic $(3_2 \Succ 1_0)\otimes(1_0 \Succ 3_{-1}) = 8_2^{\Succ\Prec\Prec} = 3_{-1} \Succ 1'_0 \Prec 3_{2} \Prec 1_0$.
Symmetrically, when we consider the product of the lowest weight modules
we find $(3_2 \Prec 1_0)\otimes(1_0 \Prec 3_{-1}) = 8_2^{\Succ\Prec\Prec} = 3_{-1} \Succ 1'_0 \Prec 3_{2} \Prec 1_0$,

In addition to the orbits of $|kk>$ and $|nn>$, the antisymmetric part of the tensor product of the two quartets contains a
shifted adjoint. If $b+b'=2$ or $b+b'=0$ we obtain a $8_3$ or a $8_{-1}$ which are indecomposable. To distinguish the
case $5_3 \Succ 3_2$ from $5_3 \Prec 3_2$, one can compute $VU|\lX\lX>$ and $UV|\mu\mu>$ to see if the triplet
$2|\mu\mu>,\lX\mu+\mu\lX>,|\lX\lX>$ is irreducibly linked to the doublet above or below to from a $5$.
Consider for example the case $b+b'=0$, $b\neq 0$. Using (F.1), we have
$bV|kk> = U|\lX\lX>=b|k\lX+\lX k>$ which is the highest weight of the irreducible triplet
$|k\lX + \lX k>, |k\mu + \mu k>,|kn + nk + \lX\mu -\mu\lX>$.
  Therefore this $3'_{-1}$ triplet is common to $4^{\Prec}_0 = 1_0 \Prec 3'_{-1}$ and
to $8^{\Succ}_{-1} = 3'_{-1} \Succ 5_{-2}$. We miss the triplet $3_{-1}$ with states 
$|k\lX - \lX k>, |k\mu - \mu k>,|kn - nk + \lX\mu +\mu\lX>$ and we can check the
$U|k\lX-\lX k> = -2b |kk>$ and $V|k\lX-\lX k> = 2 |\lX\lX>$ which belong respectively to
the $1_0$ and the $5_{-2}$. Hence we have constructed a cyclic
indecomposable representation $12^{\Succ\oplus\Succ}_{-1}=3_{-1} \Succ (1_0 \oplus 5_{-2}) \Succ 3'_{-1}$.

All cases can be analyzed in the same way, the results are listed in section 3.2.

\section{Example: indecomposable atypical 2-quiver.}

Starting from the typical shifted adjoint representation of appendix E,
if we set $b=2$, the representation becomes atypical 2. The
last 3 lines and the last 3 columns can be set to zero and we only keep the
top doublet and triplet. The matrices $U$ and $V$ reduce to
\BE
U(1,2) = \begin{pmatrix} 
      0 & 0 & 2 & 0 & 0
  \cr 0 & 0 & 0 & 2 & 0

  \cr 0 & 0 & 0 & 0 & 0
  \cr 0 & 0 & 0 & 0 & 0
  \cr 0 & 0 & 0 & 0 & 0
 \end{pmatrix}
\;,\;\;
V(1,2) = \begin{pmatrix} 
      0 & 0 & 0 & 0 & 0
  \cr 0 & 0 & 0 & 0 & 0
  
  \cr 1 & 0 & 0 & 0 & 0
  \cr 0 & 1/2 & 0 & 0 &
  \cr 0 & 0 & 0 & 0 & 0
\end{pmatrix}
\EE

Alternatively, one can construct an indecomposable matrix with
coupling parameter $\alpha$ where the last 3 states form a mouse trap.
The matrices $U$ and $V$ now read
\BE
U(1,2,\alpha) = \begin{pmatrix} 
      0 & 0 & 2 & 0 & 0 & 0 & 0 & 0
  \cr 0 & 0 & 0 & 2 & 0 & 0 & 0 & 0

  \cr 0 & 0 & 0 & 0 & 0 & 0 & 0 & 0
  \cr 0 & 0 & 0 & 0 & 0 & 0 & 0 & 0
  \cr 0 & 0 & 0 & 0 & 0 & 0 & 0 & 0
  
  \cr 0 & 0 & 0 & 0 & 0 & 0 & 1 & 0
  
 \cr 0 & 0 & 0 & 0 & 0 & 0 & 0 & 0
 \cr 0 & 0 & 0 & 0 & 0 & 0 & 0 & 0
 \end{pmatrix}
\;,\;\;\\ \\
V(1,2,\alpha) = \begin{pmatrix} 
      0 & 0 & 0 & 0 & 0 & 0 & 0 & 0
  \cr 0 & 0 & 0 & 0 & 0 & 0 & 0 & 0
  
  \cr 1 & 0 & 0 & 0 & 0 & 0 & 0 & 0
  \cr 0 & 1/2 & 0 & 0 & 0 & 0 & 0 & 0
  \cr 0 & 0 & 0 & 0 & 0 & 0 & 0 & 0

  \cr 0 & -\alpha/2 & 0 & 0 & 0 & 0 & 0 & 0
  
  \cr 0 & 0 & 0 & \alpha & 0 & 1 & 0 & 0
  \cr 0 & 0 & 0 & 0 & \alpha & 0 & 0 & 0 
 
\end{pmatrix}
\EE
The even matrices are given by equation (E.1). The other odd matrices
are given by the commutators $[E,U]=W$ and $[F,V]=-X$.

\section{Example: four generations typical indecomposable.}

Consider the fundamental typical quartet $(a=0,b)$ (singlet/doublet/singlet)
which describes the quarks of the standard model if we choose $b=4/3$. 
In the notations of section 4.2, the even matrices read
\BE
h = \begin{pmatrix} 
 0 & 0 & 0 & 0 \cr 0 & 1 & 0 & 0 \cr 0 & 0 & -1 & 0 
 \cr 0 & 0 & 0 & 0 
\end{pmatrix}
\;,\;\;\;
y = \begin{pmatrix} 
 2b & 0 & 0 & 0 \cr 0 & 2b-1 & 0 & 0 \cr 0 & 0 & 2b-1 & 0 
 \cr 0 & 0 & 0 & 2b-2 
\end{pmatrix}
\;,\;\;\;
\\
\\
f  = \begin{pmatrix} 
 0 & 0 & 0 & 0 \cr 0 & 0 & 0 & 0 \cr 0 & 1 & 0 & 0 
 \cr 0 & 0 & 0 & 0
\end{pmatrix}
\;,\;\;\;
e = \begin{pmatrix} 
 0 & 0 & 0 & 0 \cr 0 & 0 & 1 & 0 \cr 0 & 0 & 0 & 0 
 \cr 0 & 0 & 0 & 0
\end{pmatrix}
\;,\;\;\;
\EE
The odd matrices are:
\BE
u = \begin{pmatrix} 
      0 & b & 0 & 0
  \cr 0 & 0 & 0 & 0
  \cr 0 & 0 & 0 & b-1 
  \cr 0 & 0 & 0 & 0 
\end{pmatrix}
\;,\;\;\;
v = \begin{pmatrix} 
      0 & 0 & 0 & 0
  \cr 1 & 0 & 0 & 0
  \cr 0 & 0 & 0 & 0 
  \cr 0 & 0 & 1 & 0
\end{pmatrix}
\\
\\
w = \begin{pmatrix} 
      0 & 0 & -b & 0
  \cr 0 & 0 & 0 & b-1
  \cr 0 & 0 & 0 & 0 
  \cr 0 & 0 & 0 & 0 
\end{pmatrix}
\;,\;\;\;
x = \begin{pmatrix} 
      0 & 0 & 0 & 0
  \cr 0 & 0 & 0 & 0
  \cr -1 & 0 & 0 & 0 
  \cr 0 & 1 & 0 & 0
\end{pmatrix}
\;,\;\;\;
\EE
We then construct the odd matrices matrices $y'=2I$ and
\BE
u' = \begin{pmatrix} 
      0 & 1 & 0 & 0
  \cr 0 & 0 & 0 & 0
  \cr 0 & 0 & 0 & 1
  \cr 0 & 0 & 0 & 0 
\end{pmatrix}
\;,\;\;\;
w' = \begin{pmatrix} 
      0 & 0 & -1 & 0
  \cr 0 & 0 & 0 & 1
  \cr 0 & 0 & 0 & 0 
  \cr 0 & 0 & 0 & 0 
\end{pmatrix}
\EE

To construct the corresponding 12 dimensional 3 generations
odd generators as in equation (61), we only need to verify the
characteristic relations

\BE
vu' + u'v = xw' + w'x = I\;.
\EE

Hence the matrices
\BE
\LX = \begin{pmatrix} 
     \lX & \lX' & 0 & 0 \cr
     0 & \lX \;\;& \lX' & 0 \cr
     0 & 0 & \lX & \lX' \cr
     0 & 0 & 0 & \lX
\end{pmatrix}
\\
\EE
form an $SU(2/1)$ indecomposable 4-generations module.
By conjugation, we obtain the equivalent matrices
\BE
P = \begin{pmatrix} 
     I & \alpha I & 0 & 0 \cr
     0 & I \;\;& \beta I & 0 \cr
     0 & 0 & I & \gamma I \cr
     0 & 0 & 0 & I
\end{pmatrix}
\;,\;\;\;\;
P\LX P^{-1} = \begin{pmatrix} 
     \lX & \alpha \lX' & 0 & 0 \cr
     0 & \lX \;\;& \beta \lX' & 0 \cr
     0 & 0 & \lX & \gamma \lX' \cr
     0 & 0 & 0 & \lX
\end{pmatrix}
\\
\EE
Letting $p=b(b-1)$ and $q=2b-a-1$, the Casimir $C_2$ and the ghost-Casimir operator $T$  read
\BE
C_2 = \chi T = \begin{pmatrix} 
     p\,I & \alpha q\, I & \alpha\beta I & 0 \cr
     0 & p\,I \;\;& \beta q\, I & \beta\gamma I\cr
     0 & 0 & p\,I & \gamma q\,I \cr
     0 & 0 & 0 &  p\,I
\end{pmatrix}
\;,
\\
\EE
finally, letting $r=b(b-a-1)(2b-a-1)/2$, $s=3(b-a)(b-1)+((a-2)^2-3)/2$, $t=3(2b-a-1)/2$, the cubic Casimir reads
\BE
C_3 = \begin{pmatrix} 
     r\,I & \alpha s\,I & \alpha\beta t\, I & \alpha\beta\gamma\,I\cr
     0 & r\,I \;\;& \beta s\, I & \beta\gamma t\,I\cr
     0 & 0 & r\,I & \gamma s\,I\cr
     0 & 0 & 0 & r\,I
\end{pmatrix}
\;.
\\
\EE
For higher number of generations, $N>4$ the north-east triangle remains zero. 
In the original article of Marcu \cite{Marcu80}, the analysis is limited to the quartet $a=0$ and to $N<=3$
With our new method,
we can generalize the construction to any $N$ starting from any irreducible typical module.
For completeness, using the notation  $d=1/3$ we give the matrix $u'$ for the case ($a=2$,
any $b$), which goes with $u(2,b)$ equation (E.3)
\BE
u'(2) = \begin{pmatrix} 
  0\;0\;0 & 1\;0\;0\;0 & 0\;0 & 0\;0\;0 \cr
  0\;0\;0 & 0\;1\;0\;0 &-1\;0 & 0\;0\;0 \cr
   0\;0\;0 & 0\;0\;1\;0 & 0\,-1 & 0\;0\;0 \cr
   0\;0\;0 & 0\;0\;0\;0 & 0\;0 & 0\;0\;0 \cr
   0\;0\;0 & 0\;0\;0\;0 &0\;0 & d\;0\;0 \cr
   0\;0\;0 & 0\;0\;0\;0 &0\;0 & 0\;d\;0 \cr
   0\;0\;0 & 0\;0\;0\;0 &0\;0 & 0\;0\;d \cr
   0\;0\;0 & 0\;0\;0\;0 &0\;0 & d\;0\;0 \cr
   0\;0\;0 & 0\;0\;0\;0 &0\;0 & 0\;d\;0 \cr
   0\;0\;0 & 0\;0\;0\;0 &0\;0 & 0\;0\;0 \cr
   0\;0\;0 & 0\;0\;0\;0 &0\;0 & 0\;0\;0 \cr
   0\;0\;0 & 0\;0\;0\;0 &0\;0 & 0\;0\;0 \cr
\end{pmatrix}
\EE

\section{Two generation mixing of a pair of $8^{{\Succ} \oplus {\Succ}}$ cycles.}

Generalizing the constructions of appendix F, of the indecomposable cycle $8^{\Succ\oplus\Succ}$,
consider the 8 dimensional matrices

\BE
H = \begin{pmatrix} 
     0 & 0 & 0 & 0 & 0 & 0 & 0 & 0
 \cr 0 & 0 & 0 & 0 & 0 & 0 & 0 & 0
 \cr 0  & 0 & 0 & 0 & 0 & 0 & 0 & 0

 \cr 0 & 0 & 0 & 1 & 0 & 0 & 0 & 0
 \cr 0 & 0 & 0 & 0 & -1 & 0 & 0 & 0
 
 \cr 0 & 0 & 0 & 0 & 0 & 1 & 0 & 0
 \cr 0 & 0 & 0 & 0 & 0 & 0 & -1 & 0
 \cr 0 & 0 & 0 & 0 & 0 & 0 & 0 & 0
 \end{pmatrix}
\;,\;\;
Y = \begin{pmatrix} 
     0 & 0 & 0 & 0 & 0 & 0 & 0 & 0
 \cr 0 & 2 & 0 & 0 & 0 & 0 & 0 & 0

 \cr 0 & 0 & -2 & 0 & 0 & 0 & 0 & 0  
 \cr 0 & 0 & 0 & 1 & 0 & 0 & 0 & 0
 \cr 0 & 0 & 0 & 0 & 1 & 0 & 0 & 0
 
 \cr 0 & 0 & 0 & 0 & 0 & -1 & 0 & 0
  \cr 0 & 0 & 0 & 0 & 0 & 0 & -1 & 0
  
 \cr 2\alpha-2\beta & 0 & 0 & 0 & 0 & 0 & 0 & -2
 \end{pmatrix}
\;,\;\;
\\
\\
\\
F = \begin{pmatrix} 
     0 & 0 & 0 & 0 & 0 & 0 & 0 & 0
 \cr 0 & 0 & 0 & 0 & 0 & 0 & 0 & 0
 \cr 0 & 0 & 0 & 0 & 0 & 0 & 0 & 0  

 \cr 0 & 0 & 0 & 0 & 0 & 0 & 0 & 0
 \cr 0 & 0 & 0 & 1 & 0 & 0 & 0 & 0
 
 \cr 0 & 0 & 0 & 0 & 0 & 0 & 0 & 0
 \cr 0 & 0 & 0 & 0 & 0 & 1 & 0 & 0

 \cr 0 & 0 & 0 & 0 & 0 & 0 & 0 & 0
 \end{pmatrix}
\;,\;\;
E = \begin{pmatrix} 
  0 & 0 & 0 & 0 & 0 & 0 & 0 & 0
 \cr 0 & 0 & 0 & 0 & 0 & 0 & 0 & 0
 \cr 0 & 0 & 0 & 0 & 0 & 0 & 0 & 0  
 \cr 0 & 0 & 0 & 0 & 1 & 0 & 0 & 0

 \cr 0 & 0 & 0 & 0 & 0 & 0 & 0 & 0
 \cr 0 & 0 & 0 & 0 & 0 & 0 & 1 & 0
 
 \cr 0 & 0 & 0 & 0 & 0 & 0 & 0 & 0
 \cr 0 & 0 & 0 & 0 & 0 & 0 & 0 & 0
\end{pmatrix}
\;,\;\;\;
\\
\\
\\
\EE

\BE
V(\alpha,\beta) = \begin{pmatrix} 
      0 & 0 & 0 & 0 & 0 & 0 & 0 & 0
  \cr 0 & 0 & 0 & 0 & 0 & 0 & 0 & 0
  \cr 0 & 0 & 0 & 0 & 0 & 0 & -1 & 0

  \cr 0 & 1& 0 & 0 & 0 & 0 & 0 & 0
  \cr 0 & 0 & 0 & 0 & 0 & 0 & 0 & 0

  \cr -\alpha & 0 & 0 & 0 & 0 & 0 & 0 & 0
  \cr 0 & 0 & 0 & 0 & 0 & 0 & 0 & 0

 \cr 0 & 0 & 0 & 0 & \beta & 0 & 0 & 0
 \end{pmatrix}
\;,\;\;
U = \begin{pmatrix} 
      0 & 0 & 0 & 0 & 0 & 0 & 0 & 0
  \cr 0 & 0 & 0 & 1 & 0 & 0 & 0 & 0
  \cr 0 & 0 & 0 & 0 & 0 & 0 & 0 & 0

  \cr 0 & 0 & 0 & 0 & 0 & 0 & 0 & 0
  \cr -1 & 0 & 0 & 0 & 0 & 0 & 0 & 0

  \cr 0 & 0 & 0 & 0 & 0 & 0 & 0 & 0
  \cr 0 & 0 & 1 & 0 & 0 & 0 & 0 & 0

  \cr 0 & 0 & 0 & 0 & 0 & -1 & 0 & 0 
 
\end{pmatrix}
\;,\;\;\;
\\
\\
\\
\Leftrightarrow \LX(\alpha,\beta) = \begin{pmatrix} 
     0 & 0 & 0 & 0 & 0 & 0 & 0
 \cr 0 & 2y & 0 & u & -w & 0 & 0 & 0

 \cr 0 & 0 & -2y & e & 0 & -x & -v & 0  
 \cr -w & v & f & y+h & 0 & 0 & 0 & 0
 \cr -u & -x & 0 & \alpha v & y-h & 0 & 0 & 0
 
 \cr -\alpha v & 0 & w & 0 & 0 & h-y & e & 0
  \cr \alpha x & 0 & u & 0 & 0 & f & -h-y & 0
  
 \cr 2(\alpha-\beta)y & 0 & 0 & \beta x & \beta v & -u & w & 0
 \end{pmatrix}
\\
\\
\\
\;,\;\;
\EE
where the order of the singlets $(\breve\omega, kk, nn, \omega)$ are given in line and column $(1,2,3,8)$
and the doublets in columns $(4,5)$ and $(6,7)$.
If $\alpha = \beta = 4$, we recover (G.2). Otherwise, $Y$ is non diagonal and directly couples
the universal donor singlet $\breve\omega$  to the universal acceptor $\omega$.

Using these free parameters, we can construct a double matrix (4.7).
\BE
\tilde\LX = \begin{pmatrix} 
  \LX(\alpha,\beta) & \LX(\alpha '',\beta '' )\cr
  0 & \LX(\alpha ',\beta ') \cr
\end{pmatrix}
\;.
\EE
where in the lower left corner $(\alpha'',\beta'')$ we have switched the sign of the lowering operators $(V,X)$
and removed the $(F,H,E)$ $SU(2)$  operators.
If the three pairs of parameters $(\alpha,\beta)$ are distinct, the resulting 16 dimensional representation is indecomposable.
The quadratic Casimir $C_2$ vanishes and the ghost Casimir $T$ has a three non zero entries linking $\breve\omega$ to $\omega$ and to $\omega'$
proving that the representation is indecomposable.

This 16 dimensional representation was first constructed in a letter to Su (Germoni, 2000,  unpublished).
As in section 4.2, our construction can be generalized to $N$ families or applied to any other indecomposable cycle, like those appearing in the
work of G\"{o}tz, Quella and Schomerus and extended as above and classified in \cite{Germoni97,Germoni98}.
\end{appendix}
\pagebreak


\end{document}